\documentclass[12pt]{article}
\usepackage{amsmath,amssymb}
\usepackage[dvipdfmx]{graphicx}
\usepackage{color}
\usepackage{comment}

\def\mydate{21 May 2014}
\def\ignore#1{{}}

\makeatletter
\@addtoreset{equation}{section}
\makeatother

\newcommand{\beeq}{\begin{equation}}
\newcommand{\eneq}{\end{equation}}
\newcommand{\beqn}{\begin{eqnarray}}
\newcommand{\eeqn}{\end{eqnarray}}

\def\mbig{\displaystyle }
\def\dd{\partial}
\def\la{\raise.16ex\hbox{$\langle$}\lower.16ex\hbox{}  }
\def\ra{\raise.16ex\hbox{$\rangle$}\lower.16ex\hbox{} }
\def\go{\rightarrow}

\def\onehalf{ \hbox{$\frac{1}{2}$} }

\def\twothird{ \hbox{$\frac{2}{3}$} }

\def\Tr{{\rm Tr \,}}
\def\tr{{\rm tr \,}}
\def\eff{{\rm eff}}
\def\min{{\rm min}}

\def\SM{{\rm SM}}
\def\EM{{\rm EM}}

\def\KK{{\rm KK}}

\def\vect{{\rm vec}}
\def\sp{{\rm sp}}

\def\psibar{ \psi \kern-.65em\raise.6em\hbox{$-$} }
\def\psibarl{ \psi \kern-.65em\raise.6em\hbox{$-$} \lower.6em\hbox{} }

\def\Psibar{ {\overline{\Psi}} }

\def\mfrac#1#2{{\mbig #1\over \mbig #2}}

\def\mystrut{\vphantom{\mfrac{1}{\strut}}}

\begin{document}

\thispagestyle{empty}

{\small \noindent \mydate    \hfill OU-HET 806, KIAS-P14007}


\vspace{3.0cm}

\baselineskip=35pt plus 1pt minus 1pt

\begin{center}
{\LARGE \bf  LHC signals}
\vskip 10pt

{\LARGE \bf 
of the SO(5)$\times $U(1) gauge-Higgs unification
}
\end{center}

\vspace{1.5cm}
\baselineskip=22pt plus 1pt minus 1pt

\begin{center}
{\bf
Shuichiro Funatsu$^*$, Hisaki Hatanaka$^\dagger$,\\
Yutaka Hosotani$^*$\\
Yuta Orikasa$^*$ and Takuya Shimotani$^*$
}

\vskip 5pt

$^*${\small \it Department of Physics, 
Osaka University, 
Toyonaka, Osaka 560-0043, 
Japan} \\

$^\dagger${\small \it School of Physics, KIAS, Seoul 130-722, Republic of Korea}\\

\end{center}


\vskip 2.cm
\baselineskip=20pt plus 1pt minus 1pt

\begin{abstract}
Signatures of the $SO(5) \times U(1)$ gauge-Higgs unification  at  LHC and future colliders are explored.   
The Kaluza-Klein (KK) mass spectra of $\gamma, Z, Z_R$ and the Higgs self-couplings 
obey universality  relations with 
the Aharonov-Bohm phase $\theta_H$ in the fifth dimension.
The current data at low energies and at LHC indicate $\theta_H <0.2$.
Couplings of quarks and leptons to KK  gauge bosons are determined.
Three neutral gauge bosons,  the first KK modes  $Z_R^{(1)}$,  $Z^{(1)}$, and $\gamma^{(1)}$,
 appear as $Z'$ bosons in dilepton events at LHC.  
For $\theta_H = 0.114$, the mass and decay width of $Z_R^{(1)}$,  $Z^{(1)}$, and $\gamma^{(1)}$
are (5.73, 482), (6.07, 342), and (6.08\,TeV, 886\,GeV), respectively.
For $\theta_H = 0.073$ their masses are 8.00\,$\sim$\,8.61\,TeV.
An excess of events in the dilepton invariant mass should  be observed  
 in the $Z'$ search at the upgraded LHC at 14$\,$TeV.
\end{abstract}



\newpage

\baselineskip=20pt plus 1pt minus 1pt
\parskip=0pt

\section{INTRODUCTION}

The discovery of the Higgs boson of a mass around 126$\,$GeV\cite{Aad:2012tfa, Chatrchyan:2012ufa} 
supports the scenario of unification of forces and symmetry breaking envisioned in the standard model (SM) of
electroweak interactions.  Experimental data so far are consistent with what the SM describes,
but more data are necessary to pin down whether or not the discovered boson is 
definitively the Higgs boson in the SM.  Other scenarios such as supersymmetric 
models\cite{SUSYreview, Djouadi:2013vqa},  little Higgs models\cite{Arkani1}-\cite{Kalyniak:2013eva}, 
composite Higgs 
models\cite{ACP}-\cite{Carena2014},
warped extra-dimension models\cite{RS1}-\cite{Agashe:2009bb},
and UED models\cite{Appelquist2000}-\cite{Servant2014}
have been proposed in anticipation of physics beyond the SM.
It is urgent to derive and predict new phenomena
which can be observed and checked in the experiments at the upgraded 14$\,$TeV LHC.

The gauge-Higgs unification is formulated in higher-dimensional 
gauge theory \cite{YH1}-\cite{Lim2013}.
The four-dimensional Higgs boson appears as a part of the extra-dimensional component 
of gauge fields, being unified with four-dimensional gauge fields such as $W$, $Z$ and $\gamma$.  
Dynamics of the Higgs boson  are governed by the gauge principle.  
Most viable  is the $SO(5) \times U(1)$ gauge-Higgs unification
in the Randall-Sundrum warped space\cite{ACP},\cite{MSW}-\cite{FHHOS2013}.  
At low energies it yields almost the same physics 
as the SM, being consistent with LHC data and others.
Higgs couplings to gauge bosons, quarks and leptons at the tree level are suppressed 
by a common factor $\cos \theta_H$, where $\theta_H$ is the Aharonov-Bohm phase
in the extra dimension\cite{HS2007}.  
All of the precision measurements\cite{ACP, AgasheContino2006}, 
the tree-unitary constraint\cite{Haba:2009hw}, 
and the $Z'$ search\cite{Alcaraz:2006mx}-\cite{CMS:2013qca} 
indicate  $\theta_H < 0.2$.  
Branching fractions of various decay modes of the Higgs boson remain nearly the same 
as in the SM, and the signal strengths of the Higgs decay modes relative to the SM are
$\sim \cos^2 \theta_H$\cite{FHHOS2013}.   
We note that though the gauge-Higgs unification model has
similarity to the composite Higgs models, it is more restrictive and has more predictive power.

To distinguish the gauge-Higgs unification from the SM  we examine the prediction of new
particles.  It has been pointed out that the first Kaluza-Klein (KK) modes of $Z$ and $\gamma$,
denoted as  $Z^{(1)}$ and $\gamma^{(1)}$, must appear around 6$\,$TeV for $\theta_H \sim 0.1$.  
In this paper we give detailed analysis of production of $Z_R^{(1)}$, $Z^{(1)}$ and $\gamma^{(1)}$ 
at the  upgraded LHC.  Here $Z_R$ is the gauge boson associated with $SU(2)_R$, 
which does not have a zero mode.
It will be shown that $Z_R^{(1)}$, $Z^{(1)}$ and $\gamma^{(1)}$ have large widths
and can be seen as $e^+ e^-$ or $\mu^+ \mu^-$ signals.  Once their masses are
determined, the value of $\theta_H$ is fixed from the universality relations, which 
leads to further prediction of the Higgs self-couplings, etc..
Many other signals of gauge-Higgs unification have been discussed
in the literature\cite{Lim2007b}-\cite{FHHOS-DM2014}.

In Sec.2 the action of the model is given.  In addition to quark-lepton multiplets in the 
vector representation of $SO(5)$, fermion multiplets 
in the spinor representation of $SO(5)$ are introduced to realize the observed unstable Higgs boson.  
In Sec.3 the effective potential $V_\eff (\theta_H)$ is evaluated and relevant parameters 
of the model are determined.   It is shown that there appear
universality relations among $\theta_H$, the KK mass scale $m_\KK$, $m_{Z_R^{(1)}}$, 
$m_{Z^{(1)}}, m_{\gamma^{(1)}}$, and Higgs cubic and quartic couplings.
In Sec.4  dilepton ($e^+ e^-, \mu^+ \mu^-$) signals at LHC in the so-called $Z'$ search
are examined.  In the $SO(5) \times U(1)$ gauge-Higgs unification 
 $Z_R^{(1)}$, $Z^{(1)}$ and $\gamma^{(1)}$  appear as $Z'$ bosons.  
Their masses are around 6\,TeV (8\,TeV) for $\theta_H=0.114$ (0.073),  
and they have large decay widths.
We show  that they must be found in the upgraded LHC at 14\,TeV.
Sec.5 is devoted to conclusions.  
In the Appendixes we summarize KK mass spectra, wave functions and gauge couplings of 
gauge fields, quark-leptons, and $SO(5)$-spinor fermions.

\section{MODEL}

The model is defined in the Randall-Sundrum warped spacetime \cite{RS1} with the metric
\begin{eqnarray}
ds^2=G_{MN}dx^Mdx^N=e^{-2\sigma(y)} \eta_{\mu\nu}dx^\mu dx^\nu+dy^2 ~,
\label{metric1}
\end{eqnarray}
where $\eta_{\mu\nu}=\text{diag}(-1,1,1,1)$, $\sigma(-y)=\sigma(y)$, $\sigma(y+2L)=\sigma(y)$ and $\sigma(y)=k|y|$ for $|y| \leq L$.
The Planck brane and TeV brane are located at $y=0$ and $y=L$, respectively.
In the bulk region, $0< y< L$, the cosmological constant is given by $\Lambda= -6k^2$.
The warp factor $z_L=e^{kL}$ is large;  $z_L \gg 1$.
The KK mass scale is given by $m_{\KK}=\pi k/(z_L-1) \sim\pi kz_L^{-1}$.
In the fundamental region $0 \le y \le L$ the metric can be written, in terms of 
the conformal coordinate  $z= e^{ky}$,  as 
\beeq
ds^2 = \frac{1}{z^2} \Big(  \eta_{\mu\nu}dx^\mu dx^\nu+ \frac{dz^2}{k^2} \Big) ~.
\label{metric2}
\eneq

The gauge symmetry in the bulk region  is given by $SO(5)\times U(1)_X \times SU(3)_C$
with the  corresponding gauge fields $A_M$, $B_M$ and  $G_M$ and gauge couplings
$g_A, g_B$ and $g_C$.
Quark-lepton multiplets $\Psi_a$ are introduced in the vector representation {\bf 5} of
$SO(5)$, whereas additional fermions $\Psi_{F_i}$ are introduced in the 
spinor representation {\bf 4} of $SO(5)$\cite{HOOS, HNU, FHHOS2013}.
The $SO(5)$ gauge symmetry is partially broken to $SO(4) \simeq SU(2)_L \times SU(2)_R$ 
by orbifold boundary conditions.
On the Planck brane at $y=0$ ($z=1$) there live right-handed brane fermions 
$\hat \chi_{\alpha R}$ and brane scalar $\hat \Phi$, which are ({\bf 2},{\bf1})  and 
({\bf 1},{\bf 2}) representation of $SU(2)_L \times SU(2)_R$, respectively.  
The brane interactions are manifestly gauge-invariant under $SO(4) \times U(1)_X\times SU(3)_C$.
The brane scalar $\hat \Phi$ spontaneously breaks $SU(2)_R \times U(1)_X$ to $U(1)_Y$
by $\la \hat \Phi \ra \gg m_\KK$, which, in turn, induces mixing among $\Psi_a$ and 
$\hat \chi_{\alpha R}$ and makes all exotic fermions  acquire masses of $O(m_\KK)$.  
The resultant theory at low energies ($< 1\,$TeV) has the SM gauge symmetry
$SU(2)_L \times U(1)_Y \times SU(3)_C$ with the SM matter content.
All $SO(4) \times U(1)_X$ anomalies are cancelled.  Finally the $SU(2)_L \times U(1)_Y$
gauge symmetry is dynamically broken to $U(1)_\EM$ by the Hosotani mechanism.

The bulk part of the action is given by 
\beqn
&&\hskip -1.cm
S_{\text{bulk}}=\int d^5x\sqrt{-G} \, \Bigl[
-\tr  \Bigl( \frac{1}{\;4\;}F^{(A) MN} F^{(A)}_{MN}+\frac{1}{2\xi_A}
(f^{(A)}_{\text{gf}})^2+\mathcal{L}^{(A)}_{\text{gh}}\Bigr) \cr
\noalign{\kern 10pt}
&&\hskip 1.cm
-\Bigl(\frac{1}{\;4\;}F^{(B) MN} F^{(B)}_{MN}+\frac{1}{2\xi_B}
(f^{(B)}_{\text{gf}})^2+\mathcal{L}^{(B)}_{\text{gh}}\Bigr)  \cr
\noalign{\kern 10pt}
&&\hskip 1.cm
-\tr  \Bigl( \frac{1}{\;2\;}F^{(G) MN} F^{(G)}_{MN}+\frac{1}{\xi_C }
(f^{(G)}_{\text{gf}})^2+\mathcal{L}^{(G)}_{\text{gh}}\Bigr) \cr
\noalign{\kern 10pt}
&&\hskip 1.cm
+\sum_{a} \Psibar_a\mathcal{D}(c_a)\Psi_a
+ \sum_{i=1}^{n_F} \Psibar_{F_i}  \mathcal{D}(c_{F_i}) \Psi_{F_i} \Bigr],\cr
\noalign{\kern 10pt}
&&\hskip -1.cm
\mathcal{D}(c)= \Gamma^A {e_A}^M
\Big(\partial_M+\frac{1}{8}\omega_{MBC}[\Gamma^B,\Gamma^C] \cr
\noalign{\kern 10pt}
&&\hskip 1.cm
-ig_AA_M-ig_BQ_XB_M  -ig_CQ^{\text{color}}G_M)  \Big)-c\sigma'(y) ~.
\label{action1}
\eeqn
The gauge fixing and ghost terms are denoted as functionals with subscripts gf and gh, respectively. 
$F^{(A)}_{M N}=\partial_M A_N-\partial_N A_M-ig_A\bigl[A_M,A_N \bigr]$,  
$F^{(B)}_{M N}=\partial_M B_N-\partial_N B_M$,
and $F^{(G)}_{MN}= \partial_M G_N- \partial_N G_M -ig_C[G_M,G_N]$. 
The gauge fixing function is taken as 
$f_{\text{gf}}^{(A)} = z^2 \big\{  \eta^{\mu\nu} {\cal D}_\mu A_\nu
+ \xi_A k^2 z{\cal D}^c_z ( A^q_z/z) \big\} $ with a background field 
$A^c_z$ ($A_z = A^c_z + A^q_z$), $B^c_z = G^c_z=0$.
$Q^{\text{color}} =1$ for quark-multiplets and $Q^{\text{color}} =0$ otherwise.
The $SO(5)$ gauge fields $A_M$ are decomposed as
\begin{equation}
A_M=\sum^3_{a_L=1}A_M^{a_L}T^{a_L}+\sum^3_{a_R=1}A_M^{a_R}T^{a_R}
+\sum^4_{\hat{a}=1}A_M^{\hat{a}}T^{\hat{a}} ,
\end{equation}
where $T^{a_L, a_R}  (a_L , a_R = 1, 2, 3)$ and $T^{\hat{a}}  (\hat{a} = 1, 2, 3, 4)$ 
are the generators of $SO(4) \simeq SU(2)_L \times SU(2)_R$ and $SO(5)/SO(4)$, 
respectively.
The electric charge is given by
\beeq
Q_\EM =T^{3_L}+T^{3_R}+Q_X~.
\label{charge1}
\eneq
In the fermion part $\Psibar =i\Psi^\dagger \Gamma^0$ and $\Gamma^M$ 
matrices are given by
\begin{equation}\Gamma^{\mu}=\begin{pmatrix}
 &\sigma^{\mu}\\  {\bar \sigma}^{\mu}& \end{pmatrix}\;,\;\;
 \Gamma^5=\begin{pmatrix} 1& \\ &-1\end{pmatrix} , \;\;  
\sigma^{\mu}=(1,\;\vec{\sigma})\;,\;\;
{\bar \sigma}^{\mu}=(-1,\;\vec{\sigma}) ~.
\label{DiracGamma}
\end{equation}
The $c \sigma'(y)$ term in the action (\ref{action1}) gives a bulk kink mass.
The dimensionless parameter $c$ plays an important role in controlling profiles of 
fermion wave functions.

The orbifold boundary conditions at $y_0 = 0$ and $y_1 = L$ are given by
\beqn
&&\hskip -1.cm
\begin{pmatrix}A_{\mu}\\A_y\end{pmatrix} (x,y_j-y)=
P_{\vect} \begin{pmatrix}A_{\mu}\\-A_y\end{pmatrix} (x,y_j+y) P_{\vect}^{-1},  \cr
\noalign{\kern 10pt}
&&\hskip -1.cm
\begin{pmatrix}B_{\mu}\\B_y\end{pmatrix} (x,y_j-y)=
\begin{pmatrix}B_{\mu}\\-B_y\end{pmatrix}(x,y_j+y), \cr
\noalign{\kern 10pt}
&&\hskip -1.cm
\begin{pmatrix}G_{\mu}\\G_y\end{pmatrix} (x,y_j-y)=
\begin{pmatrix}G_{\mu}\\-G_y\end{pmatrix}(x,y_j+y), \cr
\noalign{\kern 10pt}
&&\hskip -1.cm
\Psi_a(x,y_j-y)=P_{\vect} \Gamma^5 \Psi_a(x,y_j+y),\cr
\noalign{\kern 10pt}
&&\hskip -1.cm
\Psi_{F_i} (x, y_j -y)=(-1)^j P_{\sp} \Gamma^5\Psi_{F_i} (x, y_j + y),  \cr
\noalign{\kern 10pt}
&&\hskip -1.cm
P_{\vect}=\text{diag} \, ( -1,-1, -1,-1, +1) , \quad 
P_{\sp} =\text{diag} \, ( +1,+1, -1,-1 ) .
\label{BC1}    
\eeqn
The $SO(5)$ symmetry is reduced to  $SO(4) \simeq SU(2)_L\times SU(2)_R$ 
by the orbifold boundary conditions. 
At this stage the four-dimensional components of the five-dimensional gauge fields have 
zero modes in $SO(4) \times U(1)_X \times SU(3)_C$, whereas the extra-dimensional
components have zero modes  in $SO(5)/SO(4)$, $A_y^{\hat a}$ or 
$A_z^{\hat a}$ ($a=1, \cdots, 4$). The latter contains the four-dimensional Higgs field, which is
a doublet both in $SU(2)_L$ and in $SU(2)_R$.
Without loss of generality one can set $\langle A^{\hat{a}}_y\rangle  \propto \delta^{a4}$ 
when the EW symmetry is spontaneously broken by the Hosotani mechanism. 
The zero modes of $A^{\hat{a}}_y$ (a = 1,2,3) are absorbed by $W$ and $Z$ bosons. 
The four-dimensional neutral Higgs field $H(x)$ is a fluctuation mode of 
the Wilson line phase $\theta_H$,
\beqn
&&\hskip -1.cm
A^{\hat{4}}_y (x,y)= \big\{ \theta_Hf_H +H(x) \big\} \tilde u_H(y)+\cdots ~, \cr
\noalign{\kern 10pt}
&&\hskip -1.cm
\exp \Big\{ \frac{i}{2} \theta_H \cdot 2 \sqrt{2}T^{\hat{4}} \Big\} 
= \exp  \bigg\{ ig_A	\int^L_0 dy  \la A_y \ra \bigg\}   ~, \cr
\noalign{\kern 10pt}
&&\hskip -1.cm
f_H=\frac{2}{g_A} \sqrt{\frac{k}{z_L^2-1}}
=\frac{2}{g_w} \sqrt{\frac{k}{L(z_L^2-1)}} ~.
\label{WilsonPhase1}
\eeqn
Here the wave function of the four-dimensional Higgs boson is given by 
$\tilde u_H(y) = [2k/(z_L^2 - 1)]^{1/2}e^{2ky}$ for $0 \leq y \leq L$ and 
$\tilde u_H(-y) = \tilde u_H(y) = \tilde u_H(y + 2L)$. $g_w = g_A/\sqrt{L}$ 
is the dimensionless 4 dimensional $SU(2)_L$ coupling.

Quark-lepton multiplets $\Psi_a$ are in the vector representation of $SO(5)$.
They are decomposed into $SO(4)$ vectors and singlets.  One $SO(4)$ vector multiplet
contains two $SU(2)_L$ doublets.  In each generation 
\beqn
&&\hskip -1.cm
\Psi_1=\left[ \begin{pmatrix} T \cr B \end{pmatrix},
\begin{pmatrix} t \cr b \end{pmatrix}, \, t' \, \right]_{2/3} ,   
\quad
\Psi_2= \left[ \begin{pmatrix} U \cr D \end{pmatrix},
\begin{pmatrix} X \cr Y \end{pmatrix}, \, b' \, \right]_{-1/3} , \cr
\noalign{\kern 10pt}
&&\hskip -1.cm
\Psi_3= \left[ \begin{pmatrix} \nu_\tau \cr \tau \end{pmatrix},
\begin{pmatrix} L_{1X} \cr L_{1Y}  \end{pmatrix}, \, \tau' \, \right]_{-1} , 
\quad
\Psi_4= \left[ \begin{pmatrix} L_{2X} \cr L_{2Y} \end{pmatrix},
\begin{pmatrix} L_{3X} \cr L_{3Y}  \end{pmatrix}, \, \nu'_\tau \, \right]_{0} , 
\label{QLmultiplet1}
\eeqn
where the subscripts denote $Q_X$.
We choose the bulk mass parameters such that $c_1=c_2$ and $c_3=c_4$ in each generation.
With the boundary condition in (\ref{BC1}), zero modes appear in
\beqn
&&\hskip -1.cm
\left[ Q_{1L}=\begin{pmatrix} T_L \cr B_L \end{pmatrix},
q_L= \begin{pmatrix} t_L \cr b_L \end{pmatrix}, \, t'_R \, \right] ,   
\left[ Q_{2L}=\begin{pmatrix} U_L \cr D_L \end{pmatrix},
Q_{3L}=\begin{pmatrix} X_L \cr Y_L \end{pmatrix}, \, b'_R \, \right] , \cr
\noalign{\kern 10pt}
&&\hskip -1.cm
 \left[ \ell_L= \begin{pmatrix} \nu_{\tau L} \cr \tau_L \end{pmatrix},
L_{1L}=\begin{pmatrix} L_{1XL} \cr L_{1YL}  \end{pmatrix}, \, \tau'_R \, \right] , 
 \left[ L_{2L}=\begin{pmatrix} L_{2XL} \cr L_{2YL} \end{pmatrix},
L_{3L}=\begin{pmatrix} L_{3XL} \cr L_{3YL}  \end{pmatrix}, \, \nu'_{\tau R} \, \right] .
\label{QLmultiplet2}
\eeqn

On the Planck brane there exist the brane scalar $\hat \Phi$ in ({\bf 1},{\bf 2}) representation 
of $SU(2)_L \times SU(2)_R$ with $Q_X= \onehalf$ and
the brane fermions  in  ({\bf 2},{\bf 1}) representation of $SU(2)_L \times SU(2)_R$.
\beqn
&&\hskip -1.cm
\hat{\chi}^q_{1R}= \begin{pmatrix} \hat{T}_R \cr  \hat{B}_R \end{pmatrix}_{7/6},
\quad \hat{\chi}_{2R}^q=  \begin{pmatrix} \hat{U}_R \cr  \hat{D}_R  \end{pmatrix}_{1/6},
\quad \hat{\chi}_{3R}^q= \begin{pmatrix} \hat{X}_{R} \cr  \hat{Y}_{R}  \end{pmatrix}_{-5/6} , \cr
\noalign{\kern 10pt}
&&\hskip -1.cm
\hat{\chi}^l_{1R}=  \begin{pmatrix} \hat{L}_{1XR} \cr  \hat{L}_{1YR} \end{pmatrix}_{-3/2},
\quad \hat{\chi}_{2R}^l=  \begin{pmatrix} \hat{L}_{2XR} \cr \hat{L}_{2YR} \end{pmatrix}_{1/2},
\quad \hat{\chi}_{3R}^l=  \begin{pmatrix} \hat{L}_{3XR} \cr \hat{L}_{3YR} \end{pmatrix}_{-1/2},
\label{braneF1}
\eeqn
where the subscripts denote $Q_X$.  $\hat{\chi}^q_{\alpha R}$'s are $SU(3)_C$ triplets. 
With these brane fermions all four-dimensional anomalies 
in $SO(4) \times U(1)_X$ are cancelled\cite{HNU}.

The brane part of the action is given by 
\beqn
&&\hskip -1.cm
S_{\text{brane}}= \int d^5 x \sqrt{-G} \, \delta(y) \bigg\{
-(D_\mu \hat\Phi)^\dagger D^\mu \hat\Phi
 - \lambda_{\hat \Phi}(\hat \Phi^\dagger \hat \Phi-w^2)^2  \cr
\noalign{\kern 5pt}
&&\hskip -0.5cm
+ \sum_{\alpha=1}^3 \big( \hat{\chi}^{q\dagger}_{\alpha R}  \, 
i \bar{\sigma}^\mu D_\mu \hat{\chi}^q_{\alpha R}
+\hat{\chi}^{l\dagger}_{\alpha R}i\bar{\sigma}^\mu D_\mu\hat{\chi}^l_{\alpha R} \big) \cr
\noalign{\kern 10pt}
&&\hskip -0.5cm
-i\Bigl[\kappa_1^q\hat{\chi}_{1R}^{q\dagger}\check{\Psi}_{1L}\tilde{\hat\Phi}
+\tilde{\kappa}^q\hat{\chi}_{2R}^{q\dagger}\check{\Psi}_{1L}  \hat \Phi
+\kappa_2^q\hat{\chi}_{2R}^{q\dagger}\check{\Psi}_{2L}\tilde{\hat \Phi}
+\kappa_3^q\hat{\chi}_{3R}^{q\dagger}\check{\Psi}_{2L}\hat \Phi   -(\text{h.c.})\Bigr]  \cr
\noalign{\kern 10pt}
&&\hskip -0.5cm
-i\Bigl[ \tilde{\kappa}^l\hat{\chi}_{3R}^{l\dagger}\check{\Psi}_{3L}\tilde{\hat\Phi}
+\kappa^l_1\hat{\chi}_{1R}^{l\dagger}\check{\Psi}_{3L}\hat\Phi
+\kappa_2^l\hat{\chi}_{2R}^{l\dagger}\check{\Psi}_{4L}\tilde{\hat\Phi}
+\kappa_3^l\hat{\chi}_{3R}^{l\dagger}\check{\Psi}_{4L}\hat \Phi -(\text{h.c.}) \Bigr] \bigg\} , \cr
\noalign{\kern 10pt}
&&\hskip -0.5cm
D_\mu \hat\Phi= \Big( \partial_\mu-ig_A \sum^3_{a_R=1}A^{a_R}_\mu T^{a_R}
-iQ_X g_B B_\mu\Big) \hat \Phi ~, \cr
\noalign{\kern 5pt}
&&\hskip -0.5cm
D_\mu\hat \chi_{\alpha R}  =\Big(\partial_\mu -ig_A\sum^3_{a_L=1}A^{a_L}_\mu T^{a_L}
-iQ_Xg_B B_\mu -ig_C Q^{\text{color}}G_\mu \Big) \hat \chi_{\alpha R} \, , \cr
\noalign{\kern 5pt}
&&\hskip -0.5cm
\check{\Psi}_{1L} = \begin{pmatrix} T_L & t_L \cr B_L & b_L \end{pmatrix}  ~~{\rm etc.} , 
\quad 
\tilde{\hat \Phi}=i\sigma_2 \hat \Phi^* ~. 
\label{action2}
\eeqn
$\la \hat \Phi \ra  = (0, w)^t \not= 0$ breaks  $SU(2)_R \times U(1)_X$  to $U(1)_Y$.
It also induces mass mixing on the brane
\beqn
&&\hskip -1.cm
S_{\text{brane}}^{\text{mass}} = \int  d^5 x \sqrt{-G} \, \delta(y) 
\bigg\{ -\sum^3_{\alpha=1}i\mu_\alpha^q(\hat{\chi}^{q\dagger}_{\alpha R}Q_{\alpha L}
-Q_{\alpha L}^\dagger \hat{\chi}^{q}_{\alpha R})
-i\tilde{\mu}^q(\hat{\chi}^{q\dagger}_{2 R}q_{L}-q^\dagger_{ L}\hat{\chi}^{q}_{2 R}) \cr
\noalign{\kern 10pt}
&&\hskip 1.0cm
-\sum^3_{\alpha=1} i\mu_\alpha^l(\hat{\chi}^{l\dagger}_{\alpha R}L_{\alpha L}
-L_{\alpha L}^\dagger \hat{\chi}^{l}_{\alpha R})
-i\tilde{\mu}^l(\hat{\chi}^{l\dagger}_{3 R} \ell_{L}
-\ell^\dagger_{ L}\hat{\chi}^{l}_{3 R}) \bigg\}, \cr
\noalign{\kern 10pt}
&&\hskip 1.0cm
\frac{\mu^q_\alpha}{\kappa^q_\alpha} =\frac{\tilde{\mu}^l}{\tilde{\kappa}^q}
=\frac{\mu^l_\alpha}{\kappa_\alpha^l} =\frac{\tilde{\mu}^l}{\tilde{\kappa}^l} =w ~,
\end{eqnarray}
where $\mu_\alpha,\tilde{\mu}$ define brane mass parameters. 
In the present paper we assume that the brane interactions are diagonal in the generation
of quarks and leptons.  In this case all of $\mu_\alpha,\tilde{\mu} $ and $w$ can 
be taken to be real and positive without loss of generality.
As far as $\mu_\alpha,\tilde{\mu} \gg m_\KK$, only  $\tilde \mu^q / \mu^q_2$ and
$\tilde \mu^l / \mu^l_3$ become relevant at low energies.

As shown in Sec.3, the effective potential $V_\eff (\theta_H)$ is minimized 
at $\theta_H \not=0$, thereby the electroweak symmetry breaking taking place.
The gauge fields are expanded in KK towers.  
In particular, four-dimensional components of the $SO(5) \times U(1)_X$
gauge fields are expanded, in the twisted gauge,  as
\beeq
A_\mu (x,z)  + \frac{g_B}{g_A} B_\mu (x,z)   T_B 
=  \hat{W}_\mu^- + \hat{W}_\mu^+ + \hat{Z}_\mu + \hat{A}_\mu^\gamma 
+ \hat{W}_{R\mu}^-
+ \hat{W}_{R\mu}^+ + \hat{Z}_{R\mu} + \hat{A}_\mu^{\hat{4}}~.
\label{gaugeKK1}
\eneq
Here we have introduced $T_B$ such that $\Tr  {T_B}^2 = 1$, $\Tr T_B T^\alpha=0$ and 
$\Tr T^\alpha T^\beta = \delta^{\alpha \beta}$ where $T^\alpha$'s are generators of 
$SO(5)$ in the tensorial representation.
The $\hat{W}^\pm ,   \hat{Z}$ and $  \hat{A}^\gamma $ towers contain $W^\pm, Z$ and $\gamma$.   
The other towers do not contain light modes.
Each of the $\hat{W}^+ , \hat{W}^-,   \hat{Z}$  towers splits into two KK towers at $\theta_H=0$.
In all,  (\ref{gaugeKK1}) contains 11 KK towers.  Details of wave functions of each KK tower
are tabulated in Appendix B.

The  fermion $\Psi_{F_i}$ are introduced in the spinor representation of $SO(5)$
unlike other fields in the bulk  which are  in the vector or adjoint representations\cite{FHHOS2013}.
As explained in the next section, the existence of $\Psi_{F_i}$  in addition to the other 
bulk fields leads to nontrivial dependence of the effective potential $V_\eff (\theta_H)$ on
$\theta_H$ and to the instability of the four-dimensional Higgs boson.
The boundary condition 
$\Psi_{F_i} (x, y_j -y)=(-1)^j P_{\sp} \Gamma^5\Psi_{F_i} (x, y_j + y)$ in (\ref{BC1}) 
implies that there is no zero mode for $\theta_H=0$ and 
that the lowest KK modes of $\Psi_{F_i}(x,z)$ dominantly couple to the $SU(2)_R$ gauge bosons.  
If the boundary condition $\Psi_{F_i} (x, y_j -y)=(-1)^{j +1} P_{\sp} \Gamma^5\Psi_{F_i} (x, y_j + y)$
were taken, then the lowest KK modes of $\Psi_{F_i}(x,z)$ would dominantly couple to the 
$SU(2)_L$ gauge bosons.
The lowest, neutral component of $\Psi_{F_i}$ turns out stable and becomes the dark matter
of the Universe, as will be explained  in a separate paper\cite{FHHOS-DM2014}.
For this reason the $SO(5)$-spinor fermion $\Psi_{F_i}$ is called a dark fermion.

\section{HIGGS BOSON AND THE UNIVERSALITY}

As explained in (\ref{WilsonPhase1}),  the extra-dimensional component $A_z= (kz)^{-1} A_y$ 
contains the four-dimensional Higgs field,
\beqn
&&\hskip -1.cm
A_z^{\hat 4} (x,z) = \big\{ \theta_H f_H + H(x) \big\} u_H (z) + \cdots  ~, \cr
\noalign{\kern 10pt}
&&\hskip -1.cm
u_H(z) = \sqrt{ \frac{2}{k (z_L^2 -1)} } ~  z  \quad {\rm for} ~ 1 \le z \le z_L ~.
\label{WilsonPhase2}
\eeqn
The value of $\theta_H$ is determined by the location of the global
minimum of the effective potential $V_\eff (\theta_H)$.  The Higgs boson mass is given by 
\beeq
m_H^2 = \frac{1}{f_H^2} \, \frac{d^2 V_\eff}{d \theta_H^2} \bigg|_\min ~.
\label{HiggsMass1}
\eneq
In this section we explain how the parameters of the model are determined, and show that 
universality relations appear among $\theta_H$, the KK mass $m_\KK$,
the masses of $Z^{(1)}$ and $\gamma^{(1)}$, and the Higgs self-couplings\cite{FHHOS2013}.

\subsection{$V_\eff (\theta_H)$}

Let us first consider the case in which all $SO(5)$-spinor fermions (dark fermions) $\Psi_{F_i}$ 
are degenerate at the tree level,  i.e.\ $c_{F_i} = c_F$ ($i=1, \cdots, n_F$).  
At the one-loop level only the KK towers whose
mass spectra depend on $\theta_H$ contribute to the effective potential $V_\eff (\theta_H)$.
Those spectra are given by (\ref{Wspectrum1}) for the $W$ tower, (\ref{Zspectrum1}) for 
the $Z$ tower, (\ref{Dtower1}) for the $D$ tower, (\ref{top4}) for the top quark tower,
(\ref{bottom3}) for the bottom quark tower, and (\ref{DarkF1}) for the $F$ tower or
the dark fermions.
Contributions of other quarks and leptons turn out exponentially suppressed and negligible.

The relevant parameters of the model are $k$, $z_L$, $g_A$, $g_B$, $c_t$, $\tilde \mu/\mu_2$,
$c_F$  and $n_F$, from which $V_\eff (\theta_H)$ is determined.  
Other brane mass parameters are irrelevant so long as
$\mu_\alpha, \tilde \mu, w \gg m_\KK$.  These eight parameters are chosen such that 
$m_Z$, $\alpha_w$, $\sin^2 \theta_W$, $m_t$, $m_b$, and $m_H$ take the observed 
values\cite{Beringer:1900zz}.
(To be precise, $\sin^2 \theta_W$ is determined by global fit.)  This procedure leaves
two parameters, say $z_L$ and $n_F$, free.  
The procedure is highly involved as everything must be determined at the global minimum of  
$V_\eff (\theta_H)$, which, however,  is to be found after all parameters are specified.
 In other words,  all parameters must be determined self-consistently.

First we note that with those given parameters, the one--loop effective potential is given by 
\beqn
&&\hskip -1.cm
V_\eff (\theta_H, c_t, r_t,  c_F, n_F,  k, z_L, \theta_W ; \xi)
= 2(3-\xi^2) I[Q_W] + (3-\xi^2) I[Q_Z]  + 3\xi^2I[Q_D] \cr
\noalign{\kern 10pt}
&&\hskip 5.cm
-12\{I[Q_{\text{top}}]+I[Q_{\text{bottom}}]\}-8n_FI[Q_F] ~,  \cr
\noalign{\kern 10pt}
&&\hskip -0.cm
I[Q(q;\theta_H)] =
\frac{(kz_L^{-1})^4}{(4\pi)^2}\int^\infty_0 dq\, q^3 \text{ln}\{1+Q(q;\theta_H)\} ~, \cr
\noalign{\kern 10pt}
&&\hskip -0.cm
Q_W = \cos^2\theta_W Q_Z =\onehalf Q_D
=\onehalf Q_0[q; \onehalf ]\sin^2\theta_H ~, \cr
\noalign{\kern 10pt}
&&\hskip -0.cm
Q_{\text{top}} =\frac{Q_{\text{bottom}}}{r_t} =\frac{Q_0[q;c_t]}{2(1+r_t)}\sin^2\theta_H ~, \cr
\noalign{\kern 10pt}
&&\hskip -0.cm
Q_F =Q_0[q;c_F]  \cos^2 \onehalf \theta_H ~, \cr
\noalign{\kern 10pt}
&&\hskip -0.cm
Q_0[q; c] = \cfrac{z_L}{q^2\hat{F}_{c-\frac{1}{2},c-\frac{1}{2}}(qz_L^{-1},q)\hat{F}_{c+\frac{1}{2},c+\frac{1}{2}}(qz_L^{-1},q)} ~, \cr
\noalign{\kern 10pt}
&&\hskip -0.cm
\hat{F}_{\alpha,\beta}(u,v)=I_\alpha(u) K_\beta(v)-e^{-i(\alpha-\beta)\pi}K_\alpha(u)I_\beta(v) ~,
\label{effV1}
\eeqn
where $r_t = (\tilde \mu /\mu_2)^2$ and $K_\alpha$ and $I_{\alpha}$ are modified Bessel functions.
In the following we take the 't Hooft--Feynman gauge $\xi=1$.

We adopt the following algorithm to find consistent solutions.
We fix the two parameters $z_L$ and $n_F$.
\begin{enumerate}
\parskip=0pt
\item Suppose that the minimum of $V_\eff$ is located at $\theta_H = \theta_1$.
Equation (\ref{Zspectrum1})  and $\sin^2 \theta_W$ determine $\lambda_{Z^{(0)}}$, which fixes
$k$ by the $Z$ boson mass $ m_Z  = \lambda_{Z^{(0)}} k$.
\item $c_t$ and $r_t$ are determined from  (\ref{top4}) and  (\ref{bottom3}) such that
the observed masses of the top and bottom quarks are reproduced.
\item Now $V_\eff (\theta_H)$ in (\ref{effV1}) is evaluated with $c_F$ being a parameter.
$c_F$ is determined by the condition
\beeq
\frac{dV_\eff}{d \theta_H}\bigg|_{\theta_1}=0 ~,
\label{effV2}
\eneq
which assures that the minimum of $V_\eff(\theta_H)$ is located at $\theta_1$.
\item With these parameters the Higgs boson mass $m_H$ is evaluated from (\ref{HiggsMass1}).
This gives $m_H (\theta_1)$, which, in general, differs from the observed value 
$m_H = 126\,$GeV.
\item  We vary the value $\theta_1$ and 
repeat the procedure from step 1 until we get $m_H (\theta_1) = 126\,$GeV.
\end{enumerate}
In this manner the value $\theta_H = \theta_1$ at the minimum is determined as $\theta_H (z_L, n_F)$.
All other quantities such as the mass specta of all KK towers, gauge couplings of all particles,
and Yukawa couplings of all fermions are determined as functions of $z_L$, $ n_F$.
Determined values for $\theta_H$, $m_\KK$, $m_{Z^{(1)}}$, etc.\ are tabulated in 
Table \ref{nF5degenerate} in the case of $n_F=5$.

Smaller $c_t$ and $c_F$ correspond to  heavier masses of the top quark and 
dark fermions $F^{+ (1)}$ and $F^{0 (1)}$ and give larger contributions to $V_\eff (\theta_H)$.
As $n_F$ gets larger, $c_F$ ($m_{F^{(1)}}$) becomes larger (smaller) with fixed $m_H$, 
as the contribution from each  dark fermion to $V_\eff$ becomes small.
Given $n_F$, only a limited region for $z_L$ is allowed.
For $n_F=1,2,3$ one cannot reproduce the Higgs mass 126$\,$GeV when $z_L$ becomes too
small.   When $n_F\geq 4$,  one cannot reproduce the top quark mass for $z_L<10^4$.

\begin{table}[thdp]
\caption{Parameters and masses in the case of degenerate dark fermions with $n_F=5$.
All masses and $k$ are given in units of TeV.}
\begin{center}
\begin{tabular}{|c|c|c|ccc|cccc|}
\hline
$z_L\mystrut$  & $\theta_H$ & $m_\KK$& $k$& $c_t$ & $c_F$&$m_{F^{(1)}}$& $m_{Z^{(1)}_R}$
&$ m_{Z^{(1)}}$&$m_{\gamma^{(1)}}$\\
\hline
$10^9 $& 0.473 & 2.50 & $7.97\times 10^{8}$ & 0.376 & 0.459 &   0.353 & 1.92 &1.97 &1.98\\
$10^8 $& 0.351 & 3.13 & $9.97\times 10^{7} $& 0.357 & 0.445 &   0.502 & 2.40& 2.48 &  2.48\\
$10^7 $& 0.251 & 4.06 & $1.29\times 10^{7} $& 0.330 & 0.430 &   0.735  & 3.11 & 3.24& 3.24\\
$10^6 $& 0.172 & 5.45 & $1.74\times 10^6 $& 0.292 & 0.410 & 1.11   & 4.17 & 4.37  &4.38\\
 $10^5 $& 0.114 & 7.49 & $2.38\times 10^5 $& 0.227 & 0.382 & 1.75 &   5.73 & 6.07 &6.08\\
 $ 10^4 $& 0.0730 & 10.5 & $3.33\times 10^4 $& 0.0366 &   0.333 & 2.91 & 8.00& 8.61 & 8.61\\
\hline
\end{tabular}
\end{center}
\vskip -10pt
\label{nF5degenerate}
\end{table}

Dark fermions may not be degenerate.  Suppose that $n_{F}^h$ multiplets have
the bulk mass $c_{F_i} = c_{F}^h$, and $n_{F}^l$ multiplets have $c_{F_i} = c_{F}^l$.
Small difference between $c_{F}^h$ and $c_{F}^l$ can yield a substantial difference
in masses, whereas  $V_\eff (\theta_H)$ is almost unaffected. 
For instance, when $n_F = n_{F}^h + n_{F}^l = 5$,   a difference
$c_{F}^l - c_{F}^h = 0.01 (0.03)$ leads to $m_{F_h} - m_{F_l} = 30 {\rm ~to~} 80\,{\rm GeV}$
($80 {\rm ~to~} 240\,{\rm GeV}$).
The  dark fermion masses $m_{F_h^{(1)}}$ and $m_{F_l^{(1)}}$ 
in the case of $(n_{F}^h , n_{F}^l ) = (3,2)$ and $c_{F}^l - c_{F}^h = 0.03$
are tabulated in Table \ref{nFh3nFl2}.  It is found that the numerical values 
of $m_\KK$, $k$, $c_t$, $m_{Z^{(1)}_R}$, $ m_{Z^{(1)}}$, and $m_{\gamma^{(1)}}$
are the same as those in Table \ref{nF5degenerate} to the accuracy of three digits.


\begin{table}[htdp]
\caption{Parameters and masses in the case of nondegenerate dark fermions 
with $(n_{F}^h , n_{F}^l ) = (3,2)$ and  $c_{F}^l - c_{F}^h = 0.03$. 
Masses  are given in units of TeV.
The values  of $m_\KK$, $k$, $c_t$, $ m_{Z^{(1)}}$,  $m_{Z^{(1)}_R}$, 
and $m_{\gamma^{(1)}}$ are the same in three digits as those in Table \ref{nF5degenerate}
in the degenerate case.}
\begin{center}
\begin{tabular}{|c|c|ccc|}
\hline
$z_L\mystrut$  & $\theta_H$  & $c_F^h$&$m_{F_h^{(1)}}$ &$m_{F_l^{(1)}}$ \\
\hline
  $10^9$ & 0.473  & 0.447 & 0.384 & 0.304  \\
  $10^8$ & 0.351  & 0.434& 0.540 & 0.444 \\
  $10^7$ & 0.251  & 0.418 & 0.781 & 0.663 \\
  $10^6$ & 0.172   & 0.398 & 1.17 & 1.02 \\
  $10^5$ & 0.114  & 0.370& 1.83 & 1.64 \\
  $10^4$ & 0.0730 &   0.321  & 3.01 & 2.77 \\
\hline
\end{tabular}
\end{center}
\vskip -10pt
\label{nFh3nFl2}
\end{table}

\subsection{The universality}

As described above, various quantities such as $\theta_H$, $m_\KK$, the mass spectra, 
Higgs cubic and quartic self-couplings $\lambda_3, \lambda_4$, 
and Yukawa couplings are determined as functions of $z_L$ and $ n_F$
in the case of degenerate dark fermions.  In other words they depend not only on $z_L$, 
but also on how dark fermions are introduced, which could spoil the predictability of the model.
Surprisingly it has been found in Ref.\ \cite{FHHOS2013} that universal relations are held
among $\theta_H$, $m_\KK$, $m_{Z^{(1)}_R}$, $ m_{Z^{(1)}}$, $m_{\gamma^{(1)}}$, 
$\lambda_3$, and $\lambda_4$
irrespective of $n_F$.  This property is called the universality.  
It implies that once one of these quantities is determined from experiments, then other quantities
are predicted, irrespective of the details of  the  dark fermion sector.
The mass spectrum of dark fermions, $m_{F^{(1)}}$, on the other hand, 
sensitively depends on $n_F$.

\begin{figure}[tbh]\begin{center}
{\includegraphics*[width=.6\linewidth]{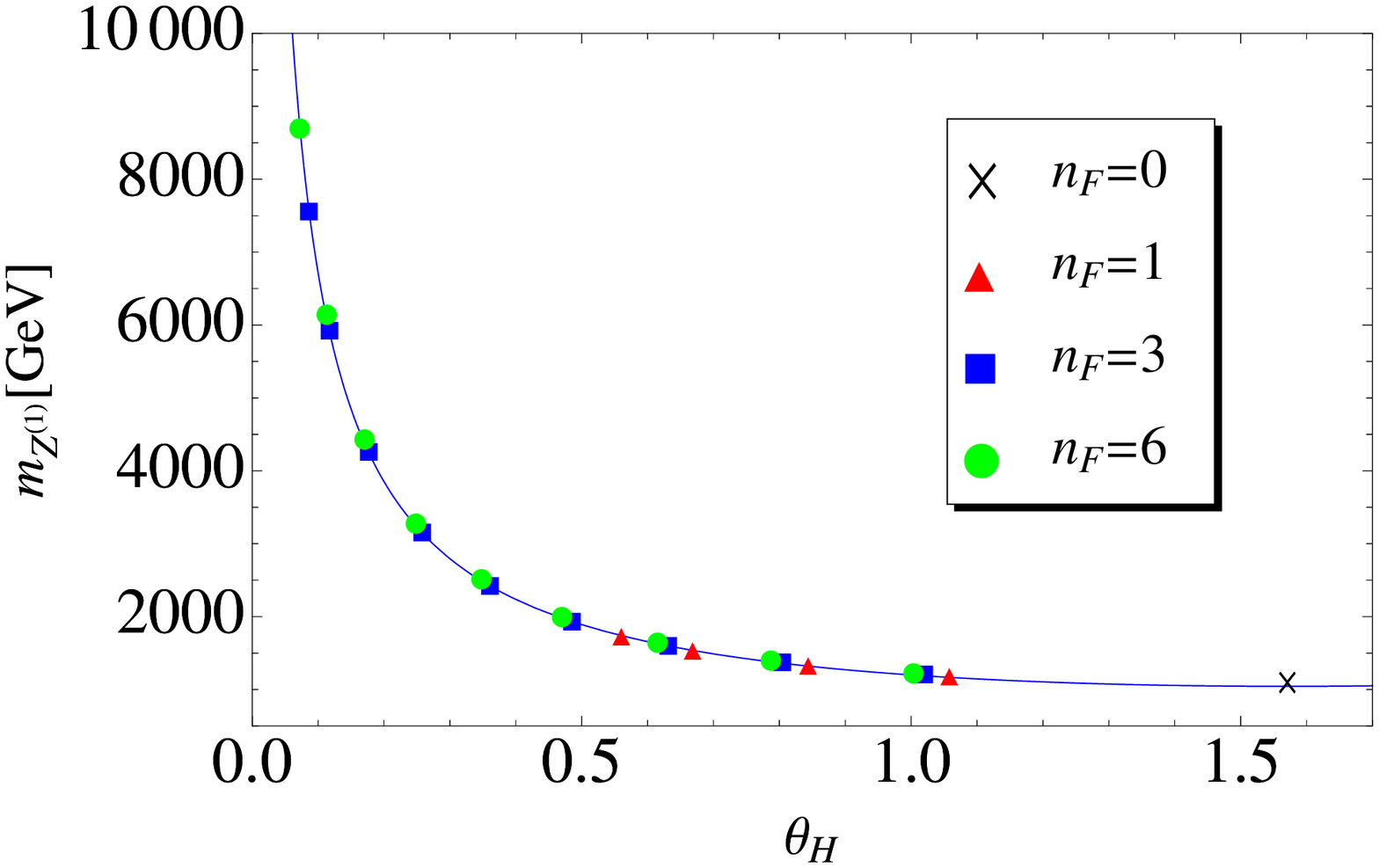}}
\caption{
$\theta_H$ vs $m_{Z^{(1)}}$ for $m_H=126 \,$GeV with $n_F$ degenerate dark fermions.
}
\label{thetaHZ1}
\end{center}
\end{figure}

It is most enlightening to express these universal relations as functions of $\theta_H$.
The masses $m_\KK$, $m_{Z^{(1)}_R}$, $ m_{Z^{(1)}}$, $m_{\gamma^{(1)}}$ are  expressed 
in the form of
\beqn
&&\hskip -1.cm
m_\KK \sim \frac{1352 \, \text{GeV}}{(\sin\theta_H)^{0.786}} ~, \cr
\noalign{\kern 10pt}
&&\hskip -1.cm
m_{Z^{(1)}_R}\sim \frac{1038 \, \text{GeV}}{(\sin\theta_H)^{0.784}} ~, \cr
\noalign{\kern 10pt}
&&\hskip -1.cm
m_{Z^{(1)}}\sim \frac{1044 \, \text{GeV}}{(\sin\theta_H)^{0.808}} ~, \cr
\noalign{\kern 10pt}
&&\hskip -1.cm
m_{\gamma^{(1)}} \sim \frac{1056 \, \text{GeV}}{(\sin\theta_H)^{0.804}} ~.
\label{universality1}
\eeqn
The relation between  $\theta_H$ and $m_{Z^{(1)}}$ is plotted in Fig.~\ref{thetaHZ1} 
for $n_F=0,1,3,6$.
One can see that the curve is universal, independent of $n_F$.
(The case of $n_F=0$ corresponds to $\theta_H = \onehalf \pi$ and the stable Higgs boson.)

\begin{figure}[htb]\begin{center}
{\includegraphics*[width=.48\linewidth]{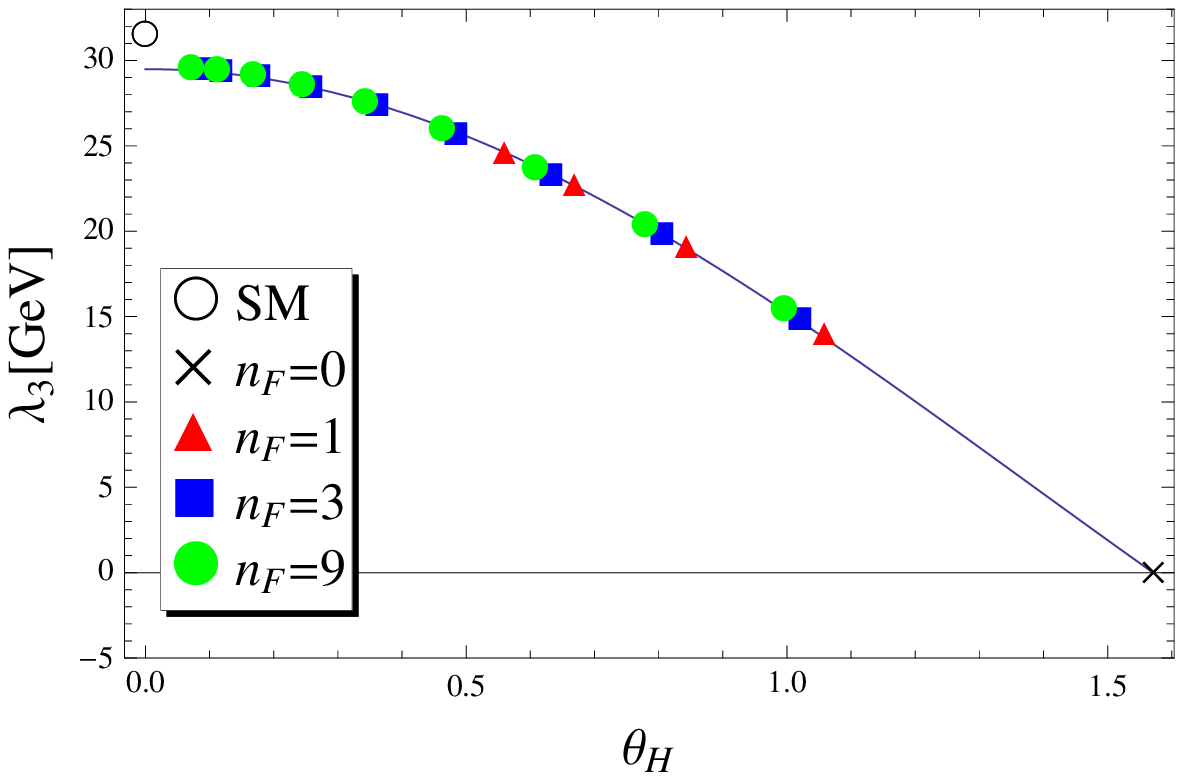}
\includegraphics*[width=.48\linewidth]{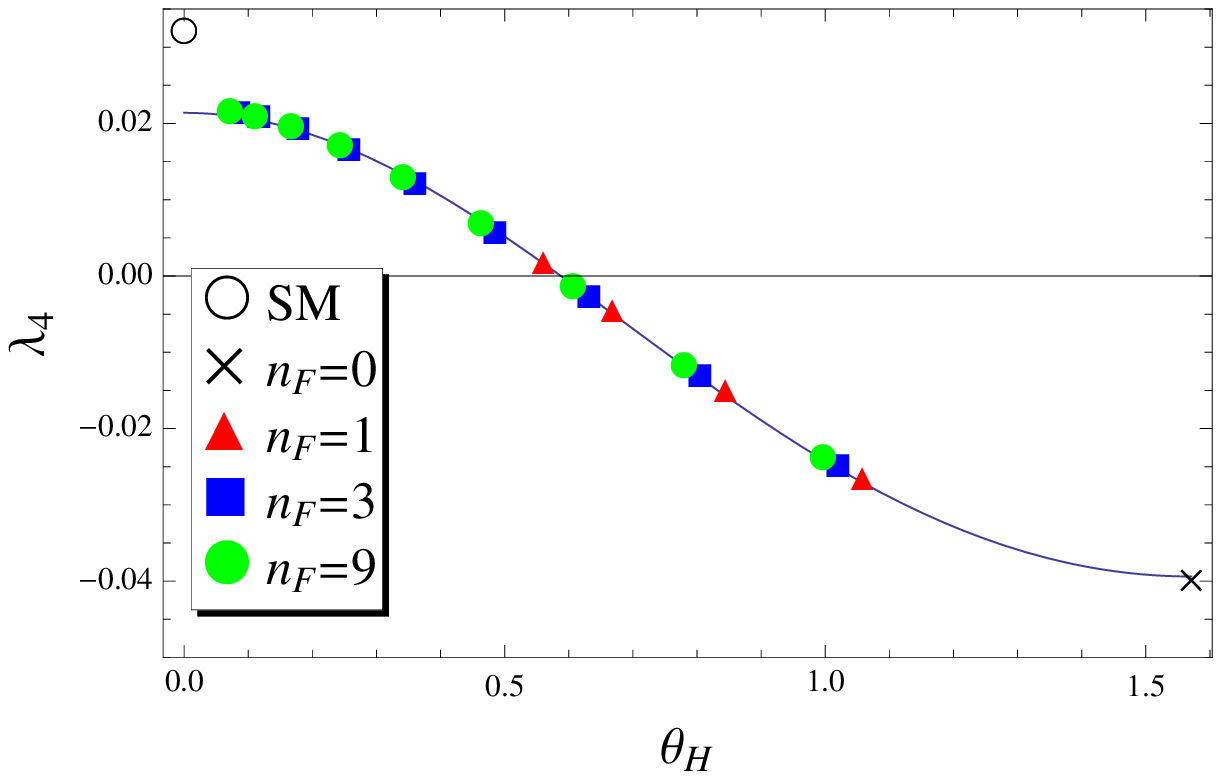}
}
\caption{
$\theta_H$ vs $\lambda_3^H$ and $\lambda_4^H$ for $m_H=126\,$GeV 
with $n_F$ degenerate dark fermions.
In the SM $\lambda_3^\SM = 31.5\,$GeV and $\lambda_4^\SM= 0.0320$.
The fitting curves are given by (\ref{HiggsCoupling}).
}
\label{thetaHlambda3}
\end{center}
\end{figure}

Similarly the Higgs cubic and quartic self-couplings,  $\lambda_3$
and $\lambda_4$ are plotted against $\theta_H$  for $n_F=0, 1,3,9$ in Fig.~\ref{thetaHlambda3}.  
The fitting curves are given by
\beqn
&&\hskip -1.cm
\lambda_3 / {\rm GeV} =  26.7 \cos \theta_H + 1.42 (1+ \cos 2\theta_H)  ~, \cr
\noalign{\kern 10pt}
&&\hskip -1.cm
\lambda_4 =  - 0.0106 + 0.0304 \cos 2 \theta_H + 0.00159 \cos 4 \theta_H ~.
\label{HiggsCoupling}
\eeqn
These numbers should be compared with $\lambda_3^\SM = 31.5\,$GeV and 
$\lambda_4^\SM= 0.0320$ in the SM.
We note that the effective potential $V_\eff(\theta_H)$ is bounded from below so that
the negative $\lambda_4$ for $\theta_H > 0.6$ does not cause the instability.
In the gauge-Higgs unification there is no instability problem in the Higgs couplings.

\begin{figure}[htb]\begin{center}
{\includegraphics*[width=.6\linewidth]{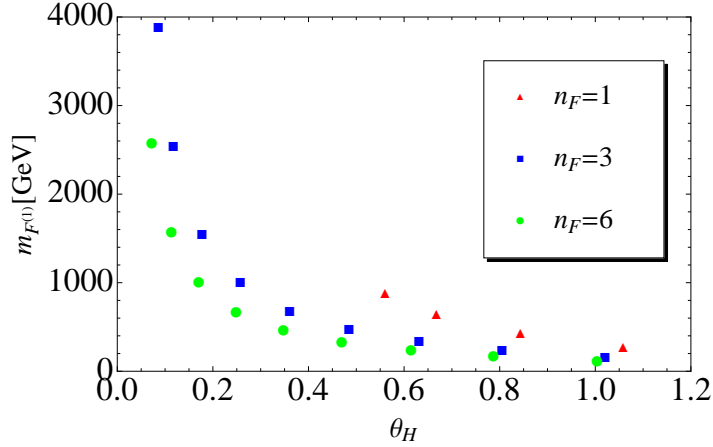}}
\caption{
$\theta_H$ vs $m_F$ for $m_H=126\,$GeV with $n_F$ degenerate dark fermions.
}
\label{thetaHmF}
\end{center}
\end{figure}

It should be noted that no universality is found in the mass spectrum of the dark fermions.  
The mass $m_{F^{(1)}}$ is plotted in  Fig.~\ref{thetaHmF} for various $n_F$.


\begin{figure}[ht]\begin{center}
{\includegraphics*[width=.6\linewidth]{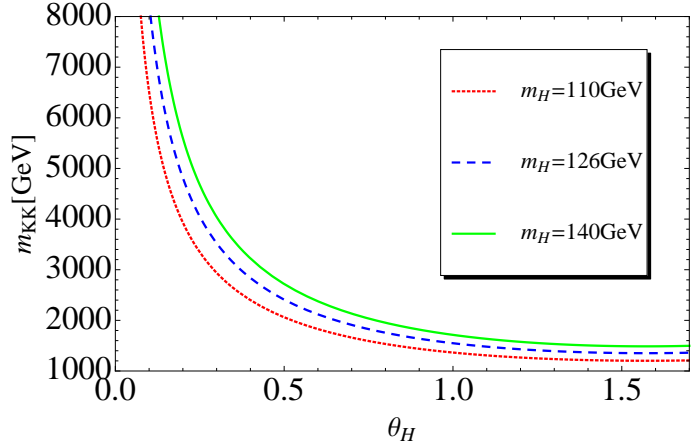}}
\caption{
$\theta_H$ vs $m_{KK}$ with various values of $m_H$.
}
\label{thetaHmKK}
\end{center}
\end{figure}

The universality relations are determined with the fixed Higgs boson mass $m_H$.
If $m_H$ were smaller or larger than the observed value, the universality relations 
would slightly change. The KK mass scale $m_\KK$ increases as $m_H$. 
The fitting curve  is parmetrized as $m_{\text{KK}}=\alpha/|\sin\theta_H|^{\beta}$ 
with given  $m_H$.  The values of $\alpha$ and $\beta$ for various $m_H$ 
are tabulated in Table \ref{thetaHmKKtable}.   We plotted $m_\KK (\theta_H)$ 
for $m_H = 110, 126, 140\,$GeV  in Fig.~\ref{thetaHmKK}.

\begin{table}[hbt]
\caption{Universality relation $m_{\text{KK}}=\alpha/|\sin\theta_H|^{\beta}$ with various value of $m_H$.}
\begin{center}
\begin{tabular}{|c|c|c|}
\hline
$m_{H}$(GeV)&$\alpha$(TeV) &$\beta$\\
\hline
110&1.20&0.733\\
120&1.30&0.766\\
126&1.35&0.786\\
130&1.39&0.800\\
140&1.49&0.820\\
\hline
\end{tabular}
\end{center}
\vskip -10pt
\label{thetaHmKKtable}
\end{table}

We stress that the universality leads to powerful predictions.  Once the value of $\theta_H$
is determined from, say, $m_{Z^{(1)}}$, many other quantities are predicted for 
experimental confirmation.  The gauge-Higgs unification scenario is very predictive.

\section{$e^+ e^-$,  $\mu^+ \mu^-$ EVENTS IN THE $Z'$ SEARCH}

One of the distinctive predictions of the $SO(5) \times U(1)$ gauge-Higgs unification
is the existence of the KK excited modes of $Z$ and $\gamma$.  Independent of the 
details of the dark fermion sector the universality predicts that 
$m_{Z^{(1)}}, m_{\gamma^{(1)}} \sim 6\,$TeV (3\,TeV)  for $\theta_H = 0.1$ (0.2)
as depicted in Fig.~\ref{thetaHZ1}.
$Z^{(1)}$ and $\gamma^{(1)}$ partially decay to $e^+ e^-$ or $\mu^+ \mu^-$, 
which should appear as clear signals in the $Z'$ search at 
LHC\cite{Basso:2010pe}-\cite{Accomando:2010fz}.
We evaluate the production and decay rates of those particles.

In our model there are four kinds of neutral gauge bosons at the TeV scale.
[See Eq.~(\ref{gaugeKK1}].
They are  the first KK mode of $Z$ boson, $Z^{(1)}$,  
the first KK mode of photon, $\gamma^{(1)}$, 
the $Z_R^{(1)}$ boson and the $A^{\hat{4}}$ boson.
Among them the $A^{\hat{4}}$ boson does not couple to SM particles
so that it escapes from detection in the $Z'$ search.
$Z^{(1)}$, $\gamma^{(1)}$, and $Z_R^{(1)}$ are the candidates for $Z^\prime$ bosons.


\subsection{Couplings and decay widths}

To evaluate the production and decay rates of $Z'$ bosons we need to know 
four-dimensional $Z'$ couplings of quarks and leptons.
They are obtained from the five-dimensional gauge interaction terms by inserting wave
functions of gauge bosons and quarks or leptons and integrating over the 
fifth-dimensional coordinate.
\ignore{
The four-dimensional gauge couplings are obtained by overlapping integrals of wave functions.
For example, the coupling between down quark and first KK mode of photon 
is included the following term.
\begin{eqnarray}
\bar{\Psi}_1\gamma^\mu T^{3_+}\Psi_1 A^{3_+}_{\mu}
\supset -\frac{\bar{d}\gamma^\mu d A_{\mu\gamma}}{\sqrt{2}}.
\end{eqnarray}
}
The couplings of the photon, $Z$ boson and $Z_R^{(1)}$ boson towers can 
be written as 
\begin{eqnarray}
 {\cal L}\supset\sum_{n,i} A^{\gamma (n)}_\mu \left[
  g_{u^iL}^{\gamma^{(n)}}\bar{u^i}^i_L\gamma^\mu u^i_L
   +g_{u^iR}^{\gamma^{(n)}}\bar{u}^i_R\gamma^\mu u^i_R
  +g_{d^iL}^{\gamma^{(n)}}\bar{d}^i_L\gamma^\mu d^i_L
   +g_{d^iR}^{\gamma^{(n)}}\bar{d}^i_R\gamma^\mu d^i_R
\right. \nonumber \\ \left.
  +g_{e^iL}^{\gamma^{(n)}}\bar{e}^i_L\gamma^\mu e^i_L
   +g_{e^iR}^{\gamma^{(n)}}\bar{e}^i_R\gamma^\mu e^i_R
 \right]
\nonumber \\
 +\sum_{n,i} Z^{(n)}_\mu
  \left[g_{u^iL}^{Z^{(n)}}\bar{u}^i_L\gamma^\mu u^i_L
   +g_{u^iR}^{Z^{(n)}}\bar{u}^i_R\gamma^\mu u^i_R
   +g_{d^iL}^{Z^{(n)}}\bar{d}^i_L\gamma^\mu d^i_L
   +g_{d^iR}^{Z^{(n)}}\bar{d}^i_R\gamma^\mu d^i_R
\right. \nonumber \\ \left.
   +g_{\nu^i L}^{Z^{(n)}}\bar{\nu}^i_L\gamma^\mu \nu^i_L
   +g_{\nu^i R}^{Z^{(n)}}\bar{\nu}^i_R\gamma^\mu \nu^i_R
   +g_{e^iL}^{Z^{(n)}}\bar{e}^i_L\gamma^\mu e^i_L
   +g_{e^iR}^{Z^{(n)}}\bar{e}^i_R\gamma^\mu e^i_R\right]
\nonumber \\
 +\sum_{n,i} Z^{(n)}_{R \mu}
  \left[g_{u^iL}^{Z_R^{(n)}}\bar{u}^i_L\gamma^\mu u^i_L
   +g_{u^iR}^{Z_R^{(n)}}\bar{u}^i_R\gamma^\mu u^i_R
   +g_{d^iL}^{Z_R^{(n)}}\bar{d}^i_L\gamma^\mu d^i_L
   +g_{d^iR}^{Z_R^{(n)}}\bar{d}^i_R\gamma^\mu d^i_R
\right. \nonumber \\ \left.
   +g_{\nu^i L}^{Z_R^{(n)}}\bar{\nu}^i_L\gamma^\mu \nu^i_L
   +g_{\nu^i R}^{Z_R^{(n)}}\bar{\nu}^i_R\gamma^\mu \nu^i_R
   +g_{e^iL}^{Z_R^{(n)}}\bar{e}^i_L\gamma^\mu e^i_L
   +g_{e^iR}^{Z_R^{(n)}}\bar{e}^i_R\gamma^\mu e^i_R\right]
\label{Zpcoupling1}
\end{eqnarray}
where the superscript $i$ denotes the generation, i.e.,\ $(u^1, u^2, u^3) = (u, c, t)$, etc.
Explicit formulas for the gauge couplings are given  in Appendix D.
The relevant couplings of the $Z^\prime$ bosons are tabulated 
 in Table \ref{Z'table0114} and Table \ref{Z'table0073}.

\begin{table}[htb]
\caption{Masses,   total decay widths and couplings 
of the $Z^\prime$ bosons to SM particles in the first generation
for $\theta_H=0.114$.
Couplings to $\mu$ are approximately the same as those to $e$.}
\begin{center}
\renewcommand{\arraystretch}{1.2}
\begin{tabular}{|l|c|c|c|c|c|c|c|c|}
\hline
   $Z'$            & $m$(TeV) & $\Gamma$(GeV) & $g_{uL}^{Z^\prime}$ 
  & $g_{dL}^{Z^\prime}$ & $g_{eL}^{Z^\prime}$ & $g_{uR}^{Z^\prime}$ 
  & $g_{dR}^{Z^\prime}$ & $g_{eR}^{Z^\prime}$ \\ \hline  
$Z$            & 0.0912              & 2.44          & 0.257 
  & $-$0.314                & $-$0.200                   & $-$0.115 
  & 0.0573                & 0.172                 \\ \hline
$Z_R^{(1)}$    & 5.73                & 482           & 0 
  & 0                     & 0                     & 0.641  
  & $-$0.321                & $-$0.978                \\ \hline
$Z^{(1)}$      & 6.07                & 342           & $-$0.0887 
  & 0.108                 & 0.0690                & $-$0.466 
  & 0.233                 & 0.711                 \\ \hline
$\gamma^{(1)}$ & 6.08                & 886           & $-$0.0724 
  & 0.0362                & 0.109                 & 0.846 
  & $-$0.423                & $-$1.29                \\ \hline
$Z^{(2)}$      & 9.14                & 1.29          & $-$0.00727 
  & 0.00889               & 0.00565               & $-$0.00548 
  & 0.00274               & 0.00856               \\ \hline
\end{tabular} 
\end{center}
\label{Z'table0114}
\end{table}

\begin{table}[htb]
\caption{Masses,   total decay widths and couplings 
of the $Z^\prime$ bosons to SM particles in the first generation
for $\theta_H=0.073$.}
\begin{center}
\renewcommand{\arraystretch}{1.2}
\begin{tabular}{|l|c|c|c|c|c|c|c|c|}
\hline
   $Z'$            & $m$(TeV) & $\Gamma$(GeV) & $g_{uL}^{Z^\prime}$ 
  & $g_{dL}^{Z^\prime}$ & $g_{eL}^{Z^\prime}$ & $g_{uR}^{Z^\prime}$ 
  & $g_{dR}^{Z^\prime}$ & $g_{eR}^{Z^\prime}$ \\ \hline  
$Z_R^{(1)}$    & 8.00                & 553           & 0 
  & 0                     & 0                     & 0.588  
  & $-$0.294                & $-$0.896                \\ \hline
$Z^{(1)}$      & 8.61                & 494           & $-$0.100 
  & 0.123                 & 0.0780                & $-$0.426 
  & 0.213                 & 0.650                 \\ \hline
$\gamma^{(1)}$ & 8.61                & 1.04$\times10^3$& $-$0.0817 
  & 0.0408                & 0.123                 & 0.775 
  & $-$0.388                & $-$1.18                \\ \hline
\end{tabular} 
\end{center}
\label{Z'table0073}
\end{table}


The decay width of the $Z^\prime$ boson is given by 
\begin{equation}
 \Gamma_{Z^\prime}=\sum_{i}\frac{m_{Z^\prime}}{12\pi}
  \left(\frac{\left(g_{iL}^{Z^\prime}\right)^2+\left(g_{iR}^{Z^\prime}\right)^2}{2}
  +2g_{iL}^{Z^\prime}g_{iR}^{Z^\prime}\frac{m_i^2}{m_{Z^\prime}^2}
\right)\sqrt{1-\frac{4m_i^2}{m_{Z^\prime}^2}} ~. 
\label{width1}
\end{equation}
Here $i$ runs over all fermions including SM fermions and dark fermions. 
The contribution of its decay to $W^+ W^-$  is very small and can be neglected\cite{HTU2}.
The evaluated $\Gamma_{Z'}$ for $\theta_H=0.114$ is summarized in Table \ref{Z'table0114}.
It is seen that all of $Z_R^{(1)}$, $Z^{(1)}$, and $\gamma^{(1)}$ have large decay widths
 ($300 \sim 900$\,GeV) in quite contrast to the narrow width of the $Z$ boson.
 It is mainly due to the large  couplings of right-handed quarks and leptons.

\subsection{Production at LHC}

In our study, we calculate the dilepton production cross sections 
through the $Z^\prime$ boson exchange together with the SM processes 
mediated by the $Z$ boson and photon. 
The dependence of the cross section on the final state 
dilepton invariant mass $M_{\ell\ell}$ is described as
\begin{eqnarray}
 \frac{d \sigma (pp \to \ell^+ \ell^- X) }
 {d M_{\ell\ell}}
 &=&  \sum_{q}
 \int^1_{-1} d \cos \theta
 \int^1_ \frac{M_{\ell\ell}^2}{E_{\rm CMS}^2} dx_1
 \frac{2 M_{\ell\ell}}{x_1 E_{\rm CMS}^2}   \nonumber \\
\noalign{\kern 10pt}
&\times & 
 f_q(x_1, M_{\ell\ell}^2)
  f_{\bar{q}} \left( \frac{M_{\ell\ell}^2}{x_1 E_{\rm CMS}^2}, M_{\ell\ell}^2
 \right)  \frac{d \sigma(\bar{q} q \to \ell^+ \ell^-) }
 {d \cos \theta},
\label{CrossLHC}
\end{eqnarray}
where $E_{\rm CMS}$ is the center-of-mass energy of the LHC and 
$f_q$'s are the parton distribution functions(PDFs) for $q$ quark.
In our numerical analysis, we employ CTEQ5M~\cite{CTEQ} for the PDFs. 
Formulas to calculate $d \sigma (\bar{q} q \to \ell^+ \ell^-)/d \cos \theta$ 
are listed in Appendix E. 

Figure~\ref{FigLHC8TeV} shows the differential cross section for $pp \to \mu^+\mu^-$ 
together with the SM cross  section mediated by the $Z$ boson and photon
for $\theta_H=0.114$ ($n_F=5$, $z_L=10^5$).
The deviation from the SM is very small below 3 TeV 
because the couplings of the $Z$ boson or photon  to SM fermions are 
almost the same as in the SM. 
For this reason it is difficult to see  the  signals of the gauge-Higgs unification  at 8 TeV LHC experiments.
In the case of $\theta_H=0.251$ ($n_F=5$, $z_L=10^7$), 
the deviation from the SM is large. 
The $Z'$ masses are around 3 TeV (See Table 1.) and the decay widths 
of $Z_R^{(1)}$ ,$Z^{(1)}$ and $\gamma^{(1)}$ are 341, 221 and 629 GeV.
The masses of $Z'$ bosons are heavier than the plot range of 
Fig.~\ref{FigLHC8TeV}.
However the decay widths of $Z'$ bosons are very wide and the deviation 
from the SM is large.
Therefore the $\theta_H=0.251$ case is excluded by the 8 TeV LHC experiments. 

\begin{figure}[ht]\begin{center}
{\includegraphics*[width=.6\linewidth]{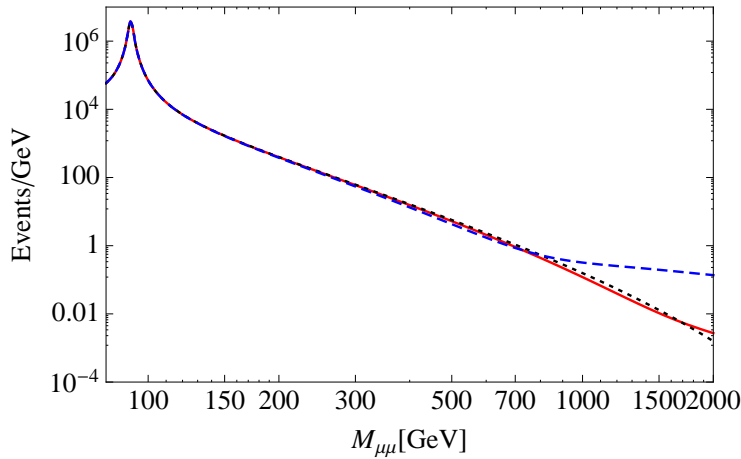}}
\caption{
The differential cross section multiplied by an integrated luminosity of 20.6 fb$^{-1}$
 for $pp \to \mu^+ \mu^- X$ 
at the 8 TeV LHC for $\theta_H=0.114$ (red solid curve) 
and for $\theta_H=0.251$ (blue dashed curve). 
The black dashed line represents the SM background.
}
\label{FigLHC8TeV}
\end{center}
\end{figure}

\begin{figure}[ht]\begin{center}
{\includegraphics*[width=.6\linewidth]{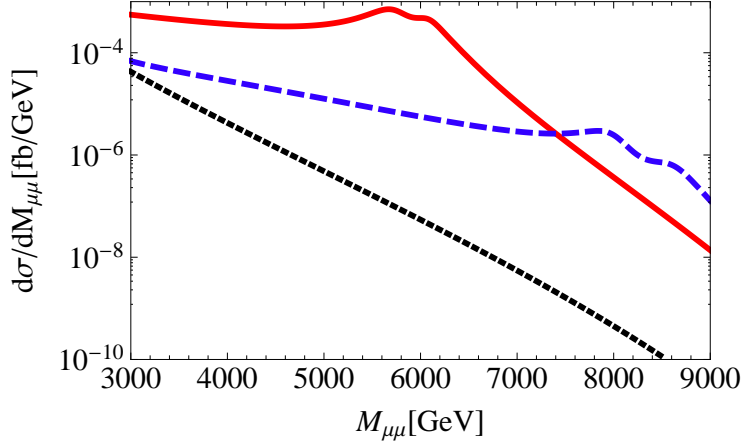}}
\caption{
The differential cross section for $pp \to \mu^+ \mu^- X$ 
at the 14 TeV LHC for $\theta_H=0.114$ (red solid curve) and 
for $\theta_H=0.073$ (blue dashed curve) .
The nearly straight line represents the SM background.
}
\label{FigLHC14TeV}
\end{center}
\end{figure}

On the other hand, at 14 TeV LHC experiments, we expect  the  signals.
Figure~\ref{FigLHC14TeV} shows the differential cross section $d \sigma/d M_{\mu \mu}$
in the range  $3\,{\rm TeV} < M_{\mu \mu} < 9\,$TeV for $\theta_H=0.114$ and 0.073.
The contributions from $Z^{(2)}$ boson and higher KK modes are negligible 
because the couplings are very small and the widths are very narrow
(see Table 4).
One sees a very large deviation from the SM, which can be detected at the upgraded LHC.

\section{CONCLUSIONS}

In the present paper we have explored LHC signals of the $SO(5) \times U(1)$ gauge-Higgs
unification, particularly dilepton events associated with the production and decay of the $Z'$ bosons
at 14\,TeV LHC.  In the $SO(5) \times U(1)$ gauge-Higgs unification the four-dimensional Higgs boson
appears as a part of the extra-dimensional component of the  $SO(5)$ gauge fields, and 
the quark or lepton  multiplets are introduced in the vector representation of $SO(5)$.   
In addition,  dark fermions are introduced in the spinor representation of $SO(5)$, 
which are vital to realize the observed unstable Higgs boson.  

The four-dimensional Higgs boson is the fluctuation mode of the Aharonov-Bohm phase $\theta_H$ in the fifth dimension.
The phase $\theta_H$, determined by the location of the global minimum of the effective 
potential $V_\eff (\theta_H)$, plays an important role in determining the  couplings 
among gauge boson,  quarks and leptons, and the Higgs boson.
It has been known that  the value $\theta_H < 0.2$ is consistent with the data
at low energies.  

The shape of $V_\eff (\theta_H)$, and therefore the location of global minimum $\theta_H$,
sensitively depends on the details of the dark fermion sector, which could spoil 
the predictability of the gauge-Higgs unification scenario.
On the contrary, we have shown that there holds the universality in the relations
among $m_\KK$, $m_{Z^{(1)}}$, $m_{\gamma^{(1)}}$, $m_{Z_R^{(1)}}$, $\lambda_3$, 
$\lambda_4$ and $\theta_H$, irrespective of the details of the dark fermion sector.
For instance, one finds that $m_{Z^{(1)}} (\theta_H) \sim 1044\,{\rm GeV}/(\sin \theta_H)^{0.804}$.
The universality implies that once the value of, say, $m_{Z^{(1)}}$ is determined from experiments,
then other quantities such as $\lambda_3$ and  $\lambda_4$ are predicted to be tested.

In the $SO(5) \times U(1)$ gauge-Higgs unification the three gauge bosons, $Z_R^{(1)}$, 
$Z^{(1)}$, and $\gamma^{(1)}$,  appear as $Z'$ bosons in dilepton events at LHC.  
It is interesting that the masses of
these bosons turn out around 6 (8\,TeV) for $\theta_H = 0.114$ ($0.073$),
which is exactly in the region  explored at the 14\,TeV LHC.
As right-handed quarks and leptons have large couplings to those $Z'$ bosons, the widths of
those bosons become large;
the decay widths of $Z_R^{(1)}$, $Z^{(1)}$ and $\gamma^{(1)}$ are 482, 342 and 886\,GeV
 (553, 494\,GeV and 1.04\,TeV) for $\theta_H=0.114$ (0.073).
Notice the relatively large ratio of ${\rm width}/{\rm mass} = 0.06 \sim 0.15$ 
in contrast to that of the $Z$ boson.
As the difference in masses of $Z^{(1)}$ and $\gamma^{(1)}$ is small,  there should appear
two peaks in dilepton events.
Due to the large widths the excess of events over those expected in the SM
should be seen in much wider  range of energies.  
For $\theta_H=0.114$, for instance,  an excess due to the broad widths of the $Z'$ resonances 
should be observed above 3\,TeV in the  dilepton invariant mass.  
The discovery of the $Z'$ bosons in the 3\,-\,9\,TeV range would give 
strong support for the gauge-Higgs unification, signaling the existence of extra dimensions.

In the present paper we have focused on the LHC signals classified in the universality class,
specifically on the $Z'$ events.  
There are other collider signals \cite{Aaltonen:2008hc}-\cite{Aad:2013cea}
such as the forward-backward asymmetry (at Tevatron)
and  the charge  asymmetry (at LHC)  in $t \bar t$ pair production
and QCD parity violation at LHC\cite{AKR}-\cite{Haba:2012bc}.  
We hope to report on these issues in the near future.

\ignore{
Further the parameters in the dark fermion sector can be fixed by direct or indirect 
observation of dark fermions.  It turns out that the neutral component of dark fermions
become the dark matter of the universe, which will be reported in a separate paper.
}

\vskip .5cm

\noindent
{\bf Acknowledgements} 

This work was supported in part  by  JSPS KAKENHI GRANTS, 
No.\ 23104009 (Y.H.\ and Y.O.),  No.\ 21244036 (Y.H.) and  No.\ 2518610 (T.S.).
H.H.\ is supported by  NRF Research Grant 2012R1A2A1A01006053 (HH) of 
the Republic of Korea.

\vskip .5cm


\begin{appendix}

\noindent

\section{BASE FUNCTIONS}\label{sec:so5generators}

Mode functions for KK towers are expressed in terms of Bessel functions.  
For gauge fields we define
\beqn
&&\hskip -1cm 
C(z;\lambda) =
      \frac{\pi}{2}\lambda z z_L F_{1,0}(\lambda z, \lambda z_L) ~, \quad
C'(z;\lambda) =
      \frac{\pi}{2}\lambda^2 z z_L F_{0,0}(\lambda z, \lambda z_L) ~ , \cr
\noalign{\kern 5pt}
&&\hskip -1cm 
S(z;\lambda) =
      -\frac{\pi}{2}\lambda z  F_{1,1}(\lambda z, \lambda z_L) ~ , \quad
S'(z;\lambda) =
      -\frac{\pi}{2}\lambda^2 z F_{0,1}(\lambda z,  \lambda z_L) ~,   \cr
\noalign{\kern 5pt}
&&\hskip -1cm 
\hat S(z;\lambda)  = \frac{C(1; \lambda)}{S(1; \lambda)} \, S(z; \lambda) ~, \cr
\noalign{\kern 5pt}
&&\hskip -1cm 
F_{\alpha,\beta}(u,v) =
    J_\alpha (u) Y_\beta(v)   -Y_\alpha(u) J_\beta(v) ~.
\label{BesselF1}
\eeqn
These functions satisfy
\beqn
&&\hskip -1cm 
C(z_L; \lambda) = z_L ~,~~ C' (z_L; \lambda) =0~, ~~ 
S(z_L; \lambda) = 0 ~,~~ S' (z_L; \lambda) =\lambda ~,  \cr
\noalign{\kern 5pt}
&&\hskip -1cm 
C S' - S C' = \lambda z ~. 
\label{BesselF2}
\eeqn
For fermions with a bulk mass parameter $c$ we define
\beqn
&&\hskip -1cm
\begin{pmatrix} C_L \cr S_L \end{pmatrix}  (z;\lambda, c)
= \pm \frac{\pi}{2} \lambda\sqrt{zz_L}
   F_{c+{1\over 2},c\mp{1\over 2}}  (\lambda z, \lambda z_L) ~, \cr
\noalign{\kern 10pt}
&&\hskip -1cm  
\begin{pmatrix} C_R \cr S_R \end{pmatrix}  (z;\lambda, c)
= \mp \frac{\pi}{2} \lambda\sqrt{zz_L}
F_{c-{1\over 2},c\pm {1\over 2}} 
 (\lambda z, \lambda z_L)~.
\label{BesselF3}
\eeqn
They satisfy
\beqn
&&\hskip -1.cm
D_+ (c)\begin{pmatrix} C_L \cr S_L \end{pmatrix} 
= \lambda \begin{pmatrix} S_R \cr C_R \end{pmatrix} ,~~
D_- (c) \begin{pmatrix} C_R \cr S_R \end{pmatrix} 
= \lambda \begin{pmatrix} S_L \cr C_L \end{pmatrix} , \cr
\noalign{\kern 10pt}
&&\hskip -1cm  
D_\pm (c)  = \pm \frac{d}{dz} + \frac{c}{z} ~,
\label{BesselF4}
\eeqn
and
\beqn
&&\hskip -1cm 
C_R=C_L =1 ~,~~  S_R=S_L=0 ~  ~~{\rm for~} z= z_L ~, \cr
\noalign{\kern 10pt}
&&\hskip -1cm  
C_L C_R - S_L S_R =1 ~, ~~
S_L(z; \lambda, -c) = - S_R(z; \lambda, c) ~.
\label{BesselF5}
\eeqn

\section{KK TOWERS OF BOSONIC FIELDS}

\subsection{Twisted gauge}

To find the spectrum and wave function of each KK mode for $\theta_H \not= 0$, it is
convenient to move to the twisted gauge in which $\la \tilde A_z \ra = 0$.
This is achieved by a gauge transformation 
$\tilde{A}_M = \Omega A_M \Omega^{-1} +1/g_A \Omega\dd_M \Omega^{-1}$ with
$\dd_z \Omega = -ig_A \Omega \la A_z \ra$;
\beqn
&&\hskip -1.cm
\Omega = \exp\left\{ig_A\theta_Hf_HT^{\hat{4}}\int^L_ydy \, u_H(y)\right\} \cr
\noalign{\kern 10pt}
&&\hskip -.6cm
= \exp \left\{i\theta_H\frac{z_L^2-z^2}{z_L^2-1}
\sqrt{2} \, T^{\hat{4}}\right\}\quad \text{for } 1\leq z\leq z_L ~.
\label{Twist1}
\eeqn
The orbifold boundary condition matrices $P_j$ at $y=y_j$ [$(y_0, y_1)=(0,L)$] 
change from $P_j$ to $\tilde P_j = \Omega(y_j -y) P_0 \Omega(y_j +y)^{-1}$.
$\Omega$ in (\ref{Twist1}) has been  chosen such that $\Omega |_{y=L}=1$
and the orbifold boundary condition at the TeV brane  remains unchanged.
On the other hand, the orbifold boundary condition matrix $P_0$ changes 
to $\tilde P_0 = \Omega(-y) P_0 \Omega(y)^{-1}$;
\beqn
&&\hskip -1.cm
\tilde P_0^{\vect} = 
\begin{pmatrix} -1&&&&\\ &-1&&&\\ &&-1&&\\
&&&-\cos2\theta_H&\sin2\theta_H\\ &&&\sin2\theta_H&\cos2\theta_H 
\end{pmatrix} , \cr
\noalign{\kern 10pt}
&&\hskip -1.cm
\tilde P_0^{\sp} = 
\begin{pmatrix} \cos \theta_H& -i \sin \theta_H \cr
i \sin \theta_H & - \cos \theta_H \end{pmatrix} \otimes I_2  ~~.
\label{twistedBC}
\eeqn

In the twisted gauge the fields satisfy free equations at the tree level, but obey
the $\theta_H$-dependent boundary condition specified by (\ref{twistedBC}).  
Wave functions of the four-dimensional components of the gauge fields are 
expressed in terms of either $C(z;\lambda)$ or $S(z;\lambda)$ in (\ref{BesselF1}), 
depending on the boundary condition (Neumann or Dirichelet) at the TeV brane.  
Wave functions of the fifth-dimensional components of the gauge fields, on the other hand, 
are  expressed in terms of either $C'(z;\lambda)$ or $S'(z;\lambda)$.
The boundary condition at the Planck brane at $z=1$ mixes fields through (\ref{twistedBC})
and determines eigenvalues $\{ \lambda_n \}$ in each KK tower.

\subsection{KK towers of $A_\mu$ and $B_\mu$}
$A_\mu(x,z)$ and $B_\mu^X(x,z)$ are expanded in KK towers.

\begin{eqnarray}
&&\hskip -1.cm
\lefteqn{
\tilde{A}_\mu(x,z) + \frac{g_B}{g_A} B_\mu (x,z) T_B 
}\nonumber\\
\noalign{\kern 10pt}
&&\hskip 0.cm
=  \hat{W}_\mu^- + \hat{W}_\mu^+ + \hat{Z}_\mu + \hat{A}_\mu^\gamma 
+ \hat{W}_{R\mu}^-
+ \hat{W}_{R\mu}^+ + \hat{Z}_{R\mu} + \hat{A}_\mu^{\hat{4}} ~,
\label{4DgaugeKK1}
\end{eqnarray}
where
\begin{eqnarray}
&&\hskip -1.cm
\hat{W}_\mu^\mp = \sum_n W_\mu^{(n)\mp}(x) 
\left\{ h_{W^{(n)}}^L \frac{T^{1_L} \mp i T^{2_L}}{\sqrt{2}} + 
h_{W^{(n)}}^R \frac{T^{1_R} \mp i T^{2_R}}{\sqrt{2}} +
\hat{h}_{W^{(n)}} \frac{T^{\hat{1}} \mp i T^{\hat{2}}}{\sqrt{2}}
\right\},
\nonumber\\
\noalign{\kern 10pt}
&&\hskip -1.cm
\hat{Z}_\mu = \sum_n Z_\mu^{(n)}(x) 
\left\{
h_{Z^{(n)}}^L T^{3_L} +
h_{Z^{(n)}}^R T^{3_R} +
\hat{h}_{Z^{(n)}} T^{\hat{3}}
+ \frac{g_B}{g_A} h_{Z^{(n)}}^B T_B 
\right\},
\nonumber\\
\noalign{\kern 10pt}
&&\hskip -1.cm
\hat{A}^{\gamma}_\mu = \sum_n A^{\gamma(n)}_\mu(x) 
\left\{
h_{\gamma^{(n)}}^L T^{3_L} +
h_{\gamma^{(n)}}^R T^{3_R} + 
\frac{g_B}{g_A} h_{\gamma^{(n)}}^B T_B 
\right\},
\nonumber\\
\noalign{\kern 10pt}
&&\hskip -1.cm
\hat{W}_{R\mu}^\mp = \sum_n W_{R\mu}^{(n)\mp}(x) 
\left\{ h_{W_R^{(n)}}^L \frac{T^{1_L} \mp i T^{2_L}}{\sqrt{2}} + 
h_{W_R^{(n)}}^R \frac{T^{1_R} \mp i T^{2_R}}{\sqrt{2}}
\right\},
\nonumber\\
\noalign{\kern 10pt}
&&\hskip -1.cm
\hat{Z}_{R\mu} = \sum_n Z_{R\mu}^{(n)}(x) 
\left\{
h_{Z_R^{(n)}}^L T^{3_L} + 
h_{Z_R^{(n)}}^R T^{3_R} +
\frac{g_B}{g_A} h_{Z_R^{(n)}}^B T_B 
\right\},
\nonumber\\
\noalign{\kern 10pt}
&&\hskip -1.cm
\hat{A}^{\hat{4}}_\mu = \sum_n A^{\hat{4}(n)}_\mu(x) 
\hat{h}_{A^{\hat{4}(n)}} T^{\hat{4}}, \cr
\noalign{\kern 10pt}
&&\hskip 0.cm
\hat{W}^{\pm}  = \frac{\hat{W}^1 \mp i \hat{W}^2}{\sqrt{2}} ~,
\quad
\hat{W}_R^{\pm} =\frac{\hat{W}_R^1 \mp i \hat{W}_R^2}{\sqrt{2}} ~.
\label{4DgaugeKK2}
\end{eqnarray}
The two gauge coupling constants are related to the weak mixing angle $\theta_W$ by
\beeq
c_\phi = \frac{g_A}{\sqrt{g_A^2 + g_B^2}} ~, ~~
s_\phi = \frac{g_B}{\sqrt{g_A^2 + g_B^2}} ~,  ~~
\cos \theta_W = \frac{1}{\sqrt{1 + s_\phi^2}} ~. 
\label{angles}
\eneq
The KK spectrum and corresponding wave functions for each tower are summarized as follows.

\paragraph{$W$ tower}
The spectrum of the $W$ tower is given by
\begin{eqnarray}
2 S(1;\lambda_{W^{(n)}}) C'(1;\lambda_{W^{(n)}}) + \lambda_{W^{(n)}} \sin^2\theta_H = 0 ~.
\label{Wspectrum1}
\end{eqnarray}
which includes the $W$ boson as the lowest mode $W = W^{(0)}$.
The mode functions are
\begin{eqnarray}
&&\hskip -1.cm 
\begin{pmatrix}
h_{W^{(n)}}^L(z) \\ h_{W^{(n)}}^R(z) \\ \hat{h}_{W^{(n)}}(z)  
\end{pmatrix}
=
\frac{1}{\sqrt{2 \, r_{W^{(n)}}}} 
\begin{pmatrix}
(1 + \cos\theta_H) C(z;\lambda_{W^{(n)}}) \\
(1 - \cos\theta_H) C(z;\lambda_{W^{(n)}}) \\
- \sqrt{2} \sin\theta_H \hat{S}(z;\lambda_{W^{(n)}})
\end{pmatrix},
\nonumber\\
\noalign{\kern 10pt}
&&\hskip -1.cm
r_{W^{(n)}} = \int_1^{z_L} \frac{dz}{kz} 
\left\{ (1 + \cos^2\theta_H) C(z; \lambda_{W^{(n)}})^2
+ \sin^2\theta_H \hat{S}(z;\lambda_{W^{(n)}})^2
\right\}.
\label{Wmode1}
\end{eqnarray}

\paragraph{$Z$ tower}
KK spectrum of the $Z$ tower is given by
\begin{eqnarray}
2 S(1;\lambda_{Z^{(n)}}) C'(1;\lambda_{Z^{(n)}}) + (1 + s_\phi^2) \lambda_{Z^{(n)}} \sin^2\theta_H = 0 ~,
\label{Zspectrum1}
\end{eqnarray}
which includes the $Z$ boson $Z=Z^{(0)}$.  
The mode functions of the $Z$ tower are
\begin{eqnarray}
&&\hskip -1.cm
\begin{pmatrix}
h_{Z^{(n)}}^L(z) \\ 
\noalign{\kern 3pt}
h_{Z^{(n)}}^R(z) \\
\noalign{\kern 3pt}
 \hat{h}_{Z^{(n)}}(z) \\
 \noalign{\kern 3pt}
 h_{Z^{(n)}}^B  
\end{pmatrix}
= \frac{1}{\sqrt{1 + s_\phi^2}}  \frac{1}{\sqrt{2 \, r_{Z^{(n)}}}} 
\begin{pmatrix}
\big\{ (1+s_\phi^2) (1+\cos\theta_H)-2s_\phi^2 \big\} C(z;\lambda_{Z^{(n)}}) \\
\noalign{\kern 3pt}
\big\{ (1+s_\phi^2) (1-\cos\theta_H)-2s_\phi^2 \big\} C(z;\lambda_{Z^{(n)}}) \\
\noalign{\kern 3pt}
 - \sqrt{2} (1+s_\phi^2)\sin\theta_H \hat{S}(z;\lambda_{Z^{(n)}}) \\
\noalign{\kern 3pt}
 - 2 s_\phi c_\phi C(z; \lambda_{Z^{(n)}})
\end{pmatrix},
\nonumber\\
\noalign{\kern 10pt}
&&\hskip -1.cm
r_{Z^{(n)}} =  \int_1^{z_L} \frac{dz}{kz} 
\biggl\{ c_\phi^2 C (z;\lambda_{Z^{(n)}})^2
\nonumber\\&&\phantom{MM} 
+ (1+s_\phi^2) [\cos^2\theta_H C(z; \lambda_{Z^{(n)}})^2
+ \sin^2\theta_H \hat{S}(z;\lambda_{Z^{(n)}})^2
\biggr\}.
\label{Zmode1}
\end{eqnarray}

\paragraph{Photon tower}
The spectrum of the photon tower is given by
\begin{eqnarray}
C'(1;\lambda_{\gamma^{(n)}}) = 0 ~,
\label{Photon1}
\end{eqnarray}
which includes a massless photon $\lambda_{\gamma^{(0)}} = 0$.
The mode functions are
\begin{eqnarray}
&&\hskip -1.cm
\begin{pmatrix}
h_{\gamma^{(n)}}^L(z) \\ h_{\gamma^{(n)}}^R(z) \\ h_{\gamma^{(n)}}^B  
\end{pmatrix}
= \frac{1}{\sqrt{1 + s_\phi^2}}  \frac{1}{\sqrt{r_{\gamma^{(n)}}}} 
\begin{pmatrix}
 s_\phi  C(z;\lambda_{\gamma^{(n)}}) \\
 s_\phi  C(z;\lambda_{\gamma^{(n)}}) \\
 c_\phi
\end{pmatrix}C(z; \lambda_{\gamma^{(n)}}),
\nonumber\\
\noalign{\kern 10pt}
&&\hskip 0.cm
r_{\gamma^{(n)}} = \int_1^{z_L} \frac{dz}{kz} 
C (z;\lambda_{\gamma^{(n)}})^2.
\label{Photon2}
\end{eqnarray}
In particular for the photon $\gamma = \gamma^{(0)}$,
\begin{eqnarray}
\begin{pmatrix}
h_{\gamma}^L(z) \\ h_{\gamma}^R(z) \\ h_{\gamma}^B  
\end{pmatrix}
&=&
\frac{1}{\sqrt{(1 + s_\phi^2)L} } 
\begin{pmatrix}
 s_\phi \\
 s_\phi \\
 c_\phi
\end{pmatrix}.
\label{Photon3}
\end{eqnarray}

\paragraph{$W_R$ tower}
The spectrum of the $W_R$ tower is given by
\begin{eqnarray}
C(1;\lambda_{W_R^{(n)}}) = 0 ~,
\label{WR1}
\end{eqnarray}
The corresponding mode functions are
\begin{eqnarray}
&&\hskip -1.cm
\begin{pmatrix}
h_{W_R^{(n)}}^L(z) \\ h_{W_R^{(n)}}^R(z) 
\end{pmatrix}
=  \frac{1}{\sqrt{2 \, r_{W_R^{(n)}}}} 
\begin{pmatrix}
+1 - \cos\theta_H   \\
-1 - \cos\theta_H
 \end{pmatrix} C(z;\lambda_{W_R^{(n)}})  ~,
\nonumber\\
\noalign{\kern 10pt}
&&\hskip 0.cm
r_{W_R^{(n)}} = \int_1^{z_L} \frac{dz}{kz}  C(z; \lambda_{W_R^{(n)}})^2 ~.
\label{WR2}
\end{eqnarray}

\paragraph{$Z_R$ tower}
The spectrum of the $Z_R$ tower is given by
\begin{eqnarray}
C(1;\lambda_{Z_R^{(n)}}) = 0 ~,
\label{ZR1}
\end{eqnarray}
turning out  identical to the $W_R$ tower spectrum, $\lambda_{Z_R^{(n)}} = \lambda_{W_R^{(n)}}$.
The corresponding mode functions are
\begin{eqnarray}
&&\hskip -1.cm
\begin{pmatrix}
h_{Z_R^{(n)}}^L(z) \\ h_{Z_R^{(n)}}^R(z)  \\ h_{Z_R^{(n)}}^B  
\end{pmatrix}
=  \frac{1}{\sqrt{1 + (1 + 2 t_\phi^2) \cos^2\theta_H} \sqrt{2 \, r_{Z_R^{(n)}}} } 
\begin{pmatrix}
-\cos\theta_H -1\\
-\cos\theta_H +1 \\
2  t_\phi  \cos\theta_H
\end{pmatrix} C(z; \lambda_{Z_R^{(n)}}) ~ ,
\nonumber\\
\noalign{\kern 10pt}
&&\hskip 0.cm
r_{Z_R^{(n)}} = \int_1^{z_L} \frac{dz}{kz} 
C (z;\lambda_{Z_R^{(n)}})^2 = r_{W_R^{(n)}},
\quad
t_\phi \equiv \frac{s_\phi}{c_\phi} ~.
\label{ZR2}
\end{eqnarray}

\paragraph{$A^{\hat{4}}$ tower}
The spectrum and wave functions of $A^{\hat{4}}$ tower are  
\begin{eqnarray}
&&\hskip -1.cm
S(1;\lambda_{A^{\hat{4}(n)}}) = 0 ~,
\\
\noalign{\kern 10pt}
&&\hskip -1.cm
h_{A^{\hat{4}(n)}}(z)
=
\frac{1}{\sqrt{r_{A^{\hat{4}(n)}}}}  S(z;\lambda_{A^{\hat{4}(n)}}),
\quad
r_{A^{\hat{4}(n)}} = \int_1^{z_L} \frac{dz}{kz} S (z;\lambda_{A^{\hat{4}(n)}})^2 ~.
\label{hatA}
\end{eqnarray}

\subsection{KK towers of $A_z$ and $B_z$}
$A_z(x,z)$ and $B_z(x,z)$ are expanded in KK towers as
\begin{eqnarray}
\tilde{A}_z (x,z) &=& \sum_{a=1}^3 \hat{G}^a + \sum_{a=1}^3 \hat{D}^a + \hat{H},
\nonumber\\
\hat{G}^a &=& \sum_{n} G^{a(n)}(x) \left\{ u_{G^{(n)}}^L T^{a_L} + u_{G^{(n)}}^R T^{a_R} \right\},
\nonumber\\
\hat{D}^a &=& \sum_{n} D^{a(n)}(x) \left\{ u_{D^{(n)}}^L T^{a_L} + u_{D^{(n)}}^R T^{a_R}
+ \hat{u}_{D^{(n)}} T^{\hat{a}} \right\},
\nonumber\\
\hat{H} &=& \sum_{n} H^{(n)}(x) u_{H^{(n)}} T^{\hat{4}},
\nonumber\\
{B}_z (x,z) &=& \sum_{n} B^{(n)}(x) u_{B^{(n)}} T_B ~.
\label{KKAz}
\end{eqnarray}

\paragraph{$G$ tower}
The $G$ tower spectrum and corresponding mode functions are given by
\begin{eqnarray}
C'(1;\lambda_{G^{(n)}}) &=& 0, \quad \lambda_{G^{(n)}} \ne 0.
\\
u_{G^{(n)}}^L = u_{G^{(n)}}^R &=& \frac{1}{\sqrt{2}} \frac{1}{\sqrt{r_{G^{(n)}}}} C'(z;\lambda_{G^{(n)}}),
\nonumber\\
r_{G^{(n)}} &=& \int_1^{z_L} \frac{kdz}{z} C'(z;\lambda_{G^{(n)}})^2.
\label{Gtower1}
\end{eqnarray}

\paragraph{$D$ tower}
The spectrum of the $D$ tower is given by
\beqn
&&\hskip -1.cm
S(1;\lambda_{D^{(n)}}) C'(1;\lambda_{D^{(n)}}) + \lambda_{D^{(n)}} \sin^2\theta_H \cr
\noalign{\kern 10pt}
&&\hskip -1.cm
= C(1;\lambda_{D^{(n)}}) S'(1;\lambda_{D^{(n)}}) - \lambda_{D^{(n)}} \cos^2\theta_H
=0 ~.
\label{Dtower1}
\eeqn
Corresponding mode functions are given by
\begin{eqnarray}
&&\hskip -1.cm
\begin{pmatrix}
h_{D^{(n)}}^L(z) \\ h_{D^{(n)}}^R(z)  \\ \hat{h}_{D^{(n)}} (z) 
\end{pmatrix}
=
\frac{1}{\sqrt{2 \, r_{D^{(n)}}}} 
\begin{pmatrix}
\cos\theta_H  C'(z;\lambda_{D^{(n)}}) \\
- \cos\theta_H C'(z;\lambda_{D^{(n)}}) \\
- \sqrt{2} \sin\theta_H \hat{S}'(z;\lambda_{D^{(n)}}) 
\end{pmatrix},
\quad
\hat{S}'(z;\lambda) = \frac{C(1;\lambda)}{S(1;\lambda)} S'(z;\lambda), 
\nonumber\\
\noalign{\kern 10pt}
&&\hskip  0.cm
r_{D^{(n)}} = \int_1^{z_L} \frac{kdz}{z} 
\left\{ 
\cos^2\theta_H C'(z;\lambda_{D^{(n)}})^2 +  
\sin^2\theta_H \hat{S}' (z;\lambda_{D^{(n)}})^2
\right\}. 
\label{Dtower2}
\end{eqnarray}

\paragraph{Higgs ($H$) tower}
The spectrum of the Higgs tower is determined by
\begin{eqnarray}
\lambda_{H^{(n)}} S (1; \lambda_{H^{(n)}}) &=& 0 ~,
\label{Higgs1}
\end{eqnarray}
which includes the zero mode $\lambda_{H^{(0)}} = 0$ for the 4D Higgs boson $H = H^{(0)}$.
The mode functions are
\begin{eqnarray}
u_{H^{(0)}}(z) &=& u_{H}(z) = \sqrt{\frac{2}{k(z_L^2-1)}} ~ z ~,
\label{Higgs2}
\end{eqnarray}
for the 4D Higgs boson, and 
\begin{eqnarray}
u_{H^{(n)}}(z) &=& \frac{1}{\sqrt{r_{H^{(n)}}}} S'(z; \lambda_{H^{(n)}}),
\quad
r_{H^{(n)}} = \int_1^{z_L} \frac{kdz}{z} S'(z;\lambda_{H^{(n)}})^2 ~,
\label{Higgs3}
\end{eqnarray}
for KK-excited states ($n \ge 1$).

\paragraph{$B$ tower}
The $B$ tower spectrum and corresponding mode functions are given by
\begin{eqnarray}
C'(1;\lambda_{B^{(n)}}) &=& 0, \quad \lambda_{B^{(n)}} \ne 0.
\\
u_{B^{(n)}} &=&  \frac{1}{\sqrt{r_{B^{(n)}}}} C'(z;\lambda_{B^{(n)}}),
\nonumber\\
r_{B^{(n)}} &=& \int_1^{z_L} \frac{kdz}{z} C'(z;\lambda_{B^{(n)}})^2.
\label{Btower1}
\end{eqnarray}


\section{WAVE FUNCTIONS OF FERMIONS}

Wave functions of KK towers of fermions are expressed in terms of
$C_L(z; \lambda, c)$ and $S_L(z; \lambda, c)$ in (\ref{BesselF3}) for left-handed components
and $C_R(z; \lambda, c)$ and $S_R(z; \lambda, c)$  for right-handed components.

\subsection{Quark-lepton towers}
Wave functions for the KK tower of an up-type quark $t$ (top)  are given by
\begin{eqnarray}
&&\hskip -1.cm
\begin{pmatrix} U_L(x,z) \\ B_L(x,z) \\ t_L(x,z) \\ t'_L(x,z) \end{pmatrix}
\supset 
\frac{\sqrt{k}z^2}{\sqrt{r_{t^{(n)}}}} 
\begin{pmatrix}
a_U^{(n)} C_L^{(2)}(z,\lambda_{t^{(n)}}) 
\\
\displaystyle
a_B^{(n)} C_L^{(1)}(z,\lambda_{t^{(n)}}) 
\\
a_t^{(n)} C_L^{(1)}(z,\lambda_{t^{(n)}})
\\
a_{t'}^{(n)} S_L^{(1)}(z,\lambda_{t^{(n)}}) 
\end{pmatrix} t^{(n)}_L(x)
\equiv
\sqrt{k}z^2 \begin{pmatrix} f^{(n)}_{U_L}(z) \\ f^{(n)}_{B_L}(z) \\ f^{(n)}_{t_L}(z) \\ f^{(n)}_{t'_L}(z) 
\end{pmatrix} t^{(n)}_L(x) ~, \cr
\noalign{\kern 10pt}
&&\hskip -1.cm
\begin{pmatrix} U_R(x,z) \\ B_R(x,z) \\ t_R(x,z) \\ t'_R(x,z) \end{pmatrix}
\supset 
\frac{\sqrt{k}z^2}{\sqrt{r_{t^{(n)}}}} 
\begin{pmatrix}
a_U^{(n)} S_R^{(2)}(z,\lambda_{t^{(n)}}) 
\\
\displaystyle
a_B^{(n)} S_R^{(1)}(z,\lambda_{t^{(n)}}) 
\\
a_t^{(n)} S_R^{(1)}(z,\lambda_{t^{(n)}})
\\
a_{t'}^{(n)} C_R^{(1)}(z,\lambda_{t^{(n)}}) 
\end{pmatrix} t^{(n)}_R(x)
\equiv 
\sqrt{k}z^2 \begin{pmatrix} f^{(n)}_{U_R}(z) \\ f^{(n)}_{B_R}(z) \\ f^{(n)}_{t_R}(z) \\ f^{(n)}_{t'_R}(z) 
\end{pmatrix} t^{(n)}_R(x) ~, \cr
\noalign{\kern 10pt}
&&\hskip -.6cm
r_{t^{(n)}} = \int_1^{z_L} dz \Big\{
a_U^{(n)2} C_L^{(2)}(z,\lambda_{t^{(n)}})^2
+ (a_B^{(n)2} + a_t^{(n)2}) C_L^{(1)}(z,\lambda_{t^{(n)}})^2 
+ a_{t'}^{(n)2} S_L^{(1)}(z,\lambda_{t^{(n)}})^2
\Big\}  \cr
\noalign{\kern 10pt}
&&\hskip -.0cm
= \int_1^{z_L} dz \Big\{
a_U^{(n)2} S_R^{(2)}(z,\lambda_{t^{(n)}})^2 
+ (a_B^{(n)2} + a_t^{(n)2}) S_R^{(1)}(z,\lambda_{t^{(n)}})^2 
+ a_{t'}^{(n)2} C_R^{(1)}(z,\lambda_{t^{(n)}})^2
\Big\}.\nonumber\\
\label{top1}
\end{eqnarray}
Here $C_L^{(i)}(z,\lambda_{t^{(n)}})=C_L(z;\lambda_{t^{(n)}},c_i)$,
$S_R^{(i)}(z,\lambda_{b^{(n)}})=S_R(z;\lambda_{b^{(n)}},c_i)$, etc., 
and other towers of $Q_\EM = \twothird e$ fermions have been suppressed.
The   common factors are given by 
\beqn
&&\hskip -1.cm
\begin{pmatrix} a^{(n)}_U \\ a^{(n)}_B \\ a^{(n)}_t \\ a^{(n)}_{t'}  \end{pmatrix}
= 
\begin{pmatrix}
-\sqrt{2} \tilde\mu^q C_L^{(1)} / \mu_2^q C_L^{(2)}  \\
\noalign{\kern 5pt}
(1- \cos \theta_H) / \sqrt{2}  \\
\noalign{\kern 5pt}
(1+ \cos \theta_H ) / \sqrt{2}  \\
\noalign{\kern 5pt}
- \sin \theta_H  C_L^{(1)} / S_L^{(1)}
\end{pmatrix}, \cr
\noalign{\kern 10pt}
&&\hskip -1.cm
C_L^{(i)} \equiv C_L(1;\lambda_{t^{(n)}},c_i) ~, ~~
S_L^{(i)} \equiv S_L(1;\lambda_{t^{(n)}},c_i) ~ , 
\label{top2}
\eeqn
where $\lambda_{t^{(n)}}$ satisfies
\beqn
&&\hskip -1.cm
(\mu_2^q )^2 C_L(1;\lambda_{t^{(n)}},c_2)   \Big\{ S_R(1;\lambda_{t^{(n)}},c_1)  
+\frac{\sin^2\theta_H}{2 S_L(1;\lambda_{t^{(n)}},c_1)} \Big\} \cr
\noalign{\kern 10pt}
&&\hskip 2.cm
 + (\tilde \mu^q)^2 C_L(1;\lambda_{t^{(n)}},c_1) S_R(1;\lambda_{t^{(n)}},c_2) = 0 ~,
\label{top3}
\eeqn
or for $c_1 = c_2 \equiv c_t$ 
\beeq
2\bigg\{ 1 + \Big( \frac{\mu_2^q}{\tilde{\mu}^q} \Big)^2 \bigg\} S_L(1;\lambda_{t^{(n)}},c_t) S_R(1;\lambda_{t^{(n)}},c_t)
+ \Big( \frac{\mu_2^q}{\tilde{\mu}^q} \Big)^2 \sin^2\theta_H = 0 ~.
 \label{top4}
\eneq

For a down-type quark $b$ (bottom) we have
\begin{eqnarray}
&&\hskip -1.cm
\begin{pmatrix} b_L(x,z) \\ X_L(x,z) \\ D_L(x,z) \\ b'_L(x,z) \end{pmatrix}
\supset 
\frac{\sqrt{k}z^2}{\sqrt{r_{b^{(n)}}}}
\begin{pmatrix}
a^{(n)}_b C_L^{(1)}(z,\lambda_{b^{(n)}}) 
\\
a^{(n)}_X C_L^{(2)}(z,\lambda_{b^{(n)}}) 
\\
a^{(n)}_D C_L^{(2)}(z,\lambda_{b^{(n)}})
\\
a^{(n)}_{b'} S_L^{(2)}(z,\lambda_{b^{(n)}})
\end{pmatrix} b^{(n)}_L(x)
\equiv
\sqrt{k}z^2
\begin{pmatrix} f^{(n)}_{b_L}(z) \\ f^{(n)}_{X_L}(z) \\ f^{(n)}_{D_L}(z) 
\\ f^{(n)}_{b'_L}(z) \end{pmatrix} b^{(n)}_L(x) ~, \cr
\noalign{\kern 10pt}
&&\hskip -1.cm
\begin{pmatrix} b_R(x,z) \\ X_R(x,z) \\ D_R(x,z) \\ b'_R(x,z) \end{pmatrix}
\supset 
\frac{\sqrt{k}z^2}{\sqrt{r_{b^{(n)}}}}
\begin{pmatrix}
a^{(n)}_b S_R^{(1)}(z,\lambda_{b^{(n)}}) 
\\
a^{(n)}_X S_R^{(2)}(z,\lambda_{b^{(n)}}) 
\\
a^{(n)}_D S_R^{(2)}(z,\lambda_{b^{(n)}})
\\
a^{(n)}_{b'} C_R^{(2)}(z,\lambda_{b^{(n)}})
\end{pmatrix} b^{(n)}_R(x)
\equiv
\sqrt{k}z^2
\begin{pmatrix} f^{(n)}_{b_R}(z) \\ f^{(n)}_{X_R}(z) \\ f^{(n)}_{D_R}(z) 
\\ f^{(n)}_{b'_R}(z) \end{pmatrix} b^{(n)}_R(x) ~, \cr
\noalign{\kern 10pt}
&&\hskip -.5cm
\begin{pmatrix} a^{(n)}_b \\ a^{(n)}_X \\ a^{(n)}_D \\ a^{(n)}_{b'} \end{pmatrix}
=
\begin{pmatrix}
-\sqrt{2}  \mu_2^q C_L^{(2)} / \tilde \mu^q C_L^{(1)}  \\
\noalign{\kern 5pt}
(1- \cos \theta_H) / \sqrt{2}  \\
\noalign{\kern 5pt}
(1+ \cos \theta_H ) / \sqrt{2}  \\
\noalign{\kern 5pt}
 \sin \theta_H  C_L^{(2)} / S_L^{(2)}
\end{pmatrix}, \cr
\noalign{\kern 10pt}
&&\hskip -.0cm
C_L^{(i)} \equiv C_L(1;\lambda_{b^{(n)}},c_i),
\quad
S_L^{(i)} \equiv S_L(1;\lambda_{b^{(n)}},c_i) ~, \cr
\noalign{\kern 10pt}
&&\hskip -1.cm
r_{b^{(n)}} = \int_1^{z_L} dz \Big\{
a^{(n)2}_b C_L^{(1)}(z,\lambda_{b^{(n)}}) ^2
+ (a^{(n)2}_X+ a^{(n)2}_D) C_L^{(2)}(z,\lambda_{b^{(n)}}) ^2
+ a^{(n)2}_{b'} S_L^{(2)}(z,\lambda_{b^{(n)}}) ^2   \Big\}   \cr
\noalign{\kern 10pt}
&&\hskip  -.6cm
= \int_1^{z_L} dz \Big\{
a^{(n)2}_b S_R^{(1)}(z,\lambda_{b^{(n)}})^2
+ (a^{(n)2}_X + a^{(n)2}_D) S_R^{(2)}(z,\lambda_{b^{(n)}})^2
+ a^{(n)2}_{b'} C_R^{(2)}(z,\lambda_{b^{(n)}})^2   \Big\}  .
\label{bottom1}
\end{eqnarray}
The spectrum is determined by
\beqn
&&\hskip -1.cm
(\tilde\mu^q)^2 C_L(1;\lambda_{b^{(n)}},c_1) 
\left[S_R(1;\lambda_{b^{(n)}},c_2) + \frac{\sin^2\theta_H}{2S_L(1;\lambda_{b^{(n)}},c_2)}\right] \cr
\noalign{\kern 10pt}
&&\hskip 2.cm
+ (\mu_2^q)^2 C_L(1;\lambda_{b^{(n)}},c_2) S_R(1;\lambda_{b^{(n)}},c_1) = 0 ~,
\label{bottom2}
\eeqn
or for $c_1=c_2=c_t$ 
\beeq
2 \bigg\{ 1 + \Big( \frac{\mu_2^q}{\tilde{\mu}^q} \Big)^2 \bigg\}
S_L(1;\lambda_{b^{(n)}},c_t)S_R(1;\lambda_{b^{(n)}},c_t) + \sin^2\theta_H = 0.
\label{bottom3}
\eneq

For a lepton mupltiplet $(\nu_\tau, \tau)$,  the wave functions are given by the following 
replacement rules;
\begin{eqnarray}
&&\hskip -1.cm
\begin{pmatrix} U \\ B \\ t \\ t' \end{pmatrix} \to
\begin{pmatrix} \nu_\tau \\ L_{2Y} \\ L_{3X} \\ \nu_\tau' \end{pmatrix},
\quad
\begin{pmatrix} b \\ D \\ X \\ b' \end{pmatrix} \to
\begin{pmatrix} L_{3Y} \\ \tau \\ L_{1Y} \\ \tau' \end{pmatrix}, \cr
\noalign{\kern 10pt}
&&\hskip -1.cm
( \tilde\mu^q , \mu_2^q ) \to (\mu_3^\ell , \tilde{\mu}^\ell) ~,
\quad
(\mu_3^q ,\mu_1^q ) \to (\mu_1^\ell, \mu_2^\ell ) ~, \cr
\noalign{\kern 10pt}
&&\hskip -1.cm
(c_1, c_2) \to (c_4,c_3) ~.
\label{lepton1}
\end{eqnarray}

\subsection{Dark fermions ($SO(5)$-spinor fermions)}

The spectrum of the KK tower of the dark fermion $\Psi_{F_i}$ is determined by
\beeq
C_L(1;\lambda_{i,n},c_{F_i}) C_R(1;\lambda_{i,n},c_{F_i}) - \sin^2 \frac{\theta_H}{2} = 0 ~.
\label{DarkF1}
\eneq
Its KK expansion is given by 
\beqn
&&\hskip -1.cm 
\Psi_{F_i ,R} (x, z) = \sqrt{k}z^2  \sum_{n=1}^\infty 
\left\{
\begin{pmatrix} f_{i,lR}^{(n)}(z) \cr 0 \cr f_{i,rR}^{(n)}(z) \cr 0 \end{pmatrix}   F^{+ (n)}_{i,R} (x)
+ \begin{pmatrix}0 \cr  f_{i,lR}^{(n)}(z) \cr 0 \cr f_{i,rR}^{(n)}(z) \end{pmatrix}   F^{0 (n)}_{i,R} (x)
\right\} ,  \cr
\noalign{\kern 10pt}
&&\hskip -1.cm 
\Psi_{F_i ,L} (x, z) = \sqrt{k}z^2  \sum_{n=1}^\infty 
\left\{
\begin{pmatrix} f_{i,lL}^{(n)}(z) \cr 0 \cr f_{i,rL}^{(n)}(z) \cr 0 \end{pmatrix}   F^{+ (n)}_{i,L} (x)
+ \begin{pmatrix}0 \cr  f_{i,lL}^{(n)}(z) \cr 0 \cr f_{i,rL}^{(n)}(z) \end{pmatrix}   F^{0 (n)}_{i,L} (x)
\right\} ,
\label{KKF1}
\eeqn
where
\begin{eqnarray}
&&\hskip -1.cm
\begin{pmatrix} f_{i, lL}^{(n)}(z) \\ f_{i, lR}^{(n)}(z) \end{pmatrix}
= \frac{i \sin \onehalf \theta_H S_L (1)}{\sqrt{r_i}}  \, 
\begin{pmatrix} C_L(z) \\ S_R (z) \end{pmatrix} 
=  \frac{\cos \onehalf \theta_H C_R(1)}{\sqrt{r_i'}}  \, 
\begin{pmatrix} C_L(z) \\ S_R (z) \end{pmatrix}, \cr
\noalign{\kern 10pt}
&&\hskip -1.cm
\begin{pmatrix} f_{i, rL}^{(n)}(z) \\ f_{i, rR}^{(n)}(z) \end{pmatrix}
=  \frac{\cos \onehalf \theta_H C_L(1)}{\sqrt{r_i}}  \, 
\begin{pmatrix} S_L(z) \\ C_R (z) \end{pmatrix} 
=  \frac{i \sin \onehalf \theta_H S_R (1)}{\sqrt{r_i'}}  \, 
\begin{pmatrix} S_L(z) \\ C_R (z) \end{pmatrix} , \cr
\noalign{\kern 10pt}
&&\hskip -.5cm
r_i = \int_1^{z_L} dz \,  \big\{ \sin^2 \onehalf \theta_H S_L(1)^2 C_L(z)^2 
+ \cos^2 \onehalf \theta_H  C_L(1)^2 S_L(z)^2 \big\}  \cr
\noalign{\kern 10pt}
&&\hskip -.1cm
= \int_1^{z_L} dz  \,  \big\{ \sin^2 \onehalf \theta_H S_L(1)^2 S_R(z)^2 
+  \cos^2 \onehalf \theta_H  C_L(1)^2 C_R(z)^2\big\} , \cr
\noalign{\kern 10pt}
&&\hskip -.5cm
r_i' = \int_1^{z_L} dz \,  \big\{ \cos^2 \onehalf \theta_H C_R(1)^2 C_L(z)^2 
+ \sin^2 \onehalf \theta_H  S_R(1)^2 S_L(z)^2 \big\}  \cr
\noalign{\kern 10pt}
&&\hskip -.1cm
= \int_1^{z_L} dz  \,  \big\{ \cos^2 \onehalf \theta_H C_R(1)^2 S_R(z)^2 
+  \sin^2 \onehalf \theta_H  S_R(1)^2 C_R(z)^2\big\} .
\label{KKF2}
\eeqn
Here $C_L(z) = C_L(z; \lambda_{i,n}, c_{F_i})$, $S_R(z) = S_R(z; \lambda_{i,n}, c_{F_i})$, etc.


\section{GAUGE COUPLINGS}

In this appendix we summarize the  couplings of quarks and leptons to the gauge bosons
and their KK excited states,  which  are necessary in evaluating dilepton events associated with the production
of the $Z'$ bosons in Sec. 4.
All four-dimensional gauge couplings of quarks and leptons are obtained from
\beeq
\int_1^{z_L} dz \sqrt{G} \, {e_m} ^\mu \sum_{a}
\bar{\tilde \Psi}_a \Gamma^m  \big( g_A \tilde A_\mu + g_B B_\mu Q_X \big) \tilde \Psi_a
\label{GaugeCouplings1}
\eneq
by inserting the wave functions of gauge bosons (\ref{4DgaugeKK2}) and 
those of fermions in Appendix C. 
Contributions coming from the interactions of the brane fermions are negligibly small,
and can be dropped below.

\subsection{$\gamma\bar{f}f$ couplings}

The couplings between the $n$ th KK photon and quarks are given by
\begin{eqnarray}
&&\hskip -1.cm
A^{\gamma(n)}_\mu(x)
\int_{1}^{z_L}dz g_A\sum_{i=1,2}\bar{\Psi}_i 
\left\{ h_{\gamma^{(n)}}^L T^{3_L} +  h_{\gamma^{(n)}}^R T^{3_R}+
\frac{g_B}{g_A} h_{\gamma^{(n)}}^B Q_X  \right\}\gamma^\mu \Psi_i  \cr
\noalign{\kern 5pt}
&&\hskip -1.cm
\supset A^{\gamma(n)}_{\mu}(x) \, g_w\sqrt{L} 
\int_{1}^{z_L}dz  \bigg[\bar{t}_L\gamma^\mu t_L(x)
\bigg\{\frac{1}{2}h_{\gamma^{(n)}}^L
    \left(f_{t_L}f_{t_L}+f_{U_L}f_{U_L}-f_{B_L}f_{B_L}\right)  \cr
\noalign{\kern 5pt}
&&\hskip -.0cm
+ \frac{1}{2}h_{\gamma^{(n)}}^R \big(f_{B_L}f_{B_L}+f_{U_L}f_{U_L}-f_{t_L}f_{t_L}\big) \cr
\noalign{\kern 5pt}
&&\hskip -.0cm
+ \frac{g_B}{3g_A} h_{\gamma^{(n)}}^B  \big( 2f_{B_L}f_{B_L}+2f_{t_L}f_{t_L}
      +2f_{t'_L}f_{t'_L}-f_{U_L}f_{U_L}  \big) \bigg\}  \cr
\noalign{\kern 5pt}
&&\hskip -.0cm
+\bar{b}_L\gamma^\mu b_L(x) \bigg\{\frac{1}{2}h_{\gamma^{(n)}}^L
    \big(f_{X_L}f_{X_L}-f_{b_L}f_{b_L}-f_{D_L}f_{D_L}\big) \cr
\noalign{\kern 5pt}
&&\hskip -.0cm
+ \frac{1}{2}h_{\gamma^{(n)}}^R \big(f_{D_L}f_{D_L}-f_{X_L}f_{X_L}-f_{b_L}f_{b_L}\big) \cr
\noalign{\kern 5pt}
&&\hskip -.0cm
+ \frac{g_B}{3g_A} h_{\gamma^{(n)}}^B  \big(2f_{b_L}f_{b_L}-f_{D_L}f_{D_L}
      -f_{b'_L}f_{b'_L}-f_{X_L}f_{X_L}\big)  \bigg\} \bigg]
 + (L\rightarrow R  ) ~.
\label{photoncoupling1}
\end{eqnarray}
$\left(L\rightarrow R \right)$ means that the wave functions of the left-handed fermions 
($f_{i L}$) are changed to those of the right-handed fermions ($f_{i R}$).
The couplings between the photon ($n=0$) and fermions are the same as those in the SM.
Similarly the couplings between the $n$ th KK photon and  leptons are given by
\begin{eqnarray}
&&\hskip -1.cm
A^{\gamma(n)}_{\mu}(x)  \, g_w\sqrt{L}  \int_{1}^{z_L}dz 
\bigg[\bar{\nu}_{\tau L}\gamma^\mu \nu_{\tau L}(x)
\bigg\{\frac{1}{2}h_{\gamma^{(n)}}^L
\left(f_{\nu_{\tau L}}f_{\nu_{\tau L}}+f_{L_{3X,L}}f_{L_{3X,L}}-f_{L_{2Y,L}}f_{L_{2Y,L}}\right) \cr
\noalign{\kern 5pt}
&&\hskip 0.cm
+ \frac{1}{2}h_{\gamma^{(n)}}^R \left(f_{\nu_{\tau L}}f_{\nu_{\tau L}}
    +f_{L_{2Y,L}}f_{L_{2Y,L}}-f_{L_{3X,L}}f_{L_{3X,L}}\right)
 - \frac{g_B}{g_A} h_{\gamma^{(n)}}^B  f_{\nu_{\tau L}}f_{\nu_{\tau L}}\bigg\} \cr
\noalign{\kern 5pt}
&&\hskip 0.cm
+\bar{\tau}_L\gamma^\mu \tau_L(x)
\bigg\{\frac{1}{2}h_{\gamma^{(n)}}^L
    \left(f_{L_{1X,L}}f_{L_{1X,L}}-f_{L_{3Y,L}}f_{L_{3Y,L}}-f_{\tau_L}f_{\tau_L}\right) \cr
\noalign{\kern 5pt}
&&\hskip 0.cm
 + \frac{1}{2}h_{\gamma^{(n)}}^R \left(f_{\tau_L}f_{\tau_L}-f_{L_{1X,L}}f_{L_{1X,L}}
-f_{L_{3Y,L}}f_{L_{3Y,L}}\right) \cr
\noalign{\kern 5pt}
&&\hskip 0.cm
- \frac{g_B}{g_A} h_{\gamma^{(n)}}^B  \left(f_{\tau_L}f_{\tau_L}+f_{\tau'_L}f_{\tau'_L}
+f_{L_{1X,L}}f_{L_{1X,L}}\right) \bigg\}\bigg] + ( L \go R) \cr
\noalign{\kern 5pt}
&&\hskip -1.cm
= A^{\gamma(n)}_{\mu}(x) \, g_w\sqrt{L}  \int_{1}^{z_L}dz  
\bar{\tau}_L\gamma^\mu \tau_L(x)
\bigg\{\frac{1}{2}h_{\gamma^{(n)}}^L
\left(f_{L_{1X,L}}f_{L_{1X,L}}-f_{L_{3Y,L}}f_{L_{3Y,L}}-f_{\tau_L}f_{\tau_L}\right) \cr
\noalign{\kern 5pt}
&&\hskip 0.cm
+ \frac{1}{2}h_{\gamma^{(n)}}^R \left(f_{\tau_L}f_{\tau_L}-f_{L_{1X,L}}f_{L_{1X,L}}
  -f_{L_{3Y,L}}f_{L_{3Y,L}}\right) \cr
\noalign{\kern 5pt}
&&\hskip 0.cm
 - \frac{g_B}{g_A} h_{\gamma^{(n)}}^B  \left(f_{\tau_L}f_{\tau_L}+f_{\tau'_L}f_{\tau'_L}
      +f_{L_{1X,L}}f_{L_{1X,L}}\right)  \bigg\}  +(L\rightarrow R) ~.
\label{photoncoupling2}
\end{eqnarray}
In the last equality, the use of  the explicit form of the  wave functions $\nu_\tau$ and 
the coupling relation (\ref{angles}) has been made.  Neutrinos do not couple to $\gamma^{(n)}$
as expected.

\subsection{$Z\bar{f}f$ couplings}

The couplings between $Z^{(n)}$ and  quarks are given by
\begin{eqnarray}
&&\hskip -1.cm
Z^{(n)}_\mu(x) \, g_w\sqrt{L} \int_{1}^{z_L}dz 
\bigg[\bar{t}_L\gamma^\mu t_L (x) \Big\{\frac{1}{2}h_{Z^{(n)}}^L
\left(f_{t_L}f_{t_L}+f_{U_L}f_{U_L}-f_{B_L}f_{B_L}\right) \cr
\noalign{\kern 5pt}
&&\hskip 0.cm
+ \frac{1}{2}h_{Z^{(n)}}^R \left(f_{B_L}f_{B_L}+f_{U_L}f_{U_L}-f_{t_L}f_{t_L}\right)
+ \hat{h}_{Z^{(n)}} \left(f_{B_L}f_{t'_L}+f_{t_L}f_{t'_L}\right)  \cr
\noalign{\kern 5pt}
&&\hskip 0.cm
+ \frac{g_B}{3g_A} h_{Z^{(n)}}^B  \left(2f_{B_L}f_{B_L}+2f_{t_L}f_{t_L}
+2f_{t'_L}f_{t'_L}-f_{U_L}f_{U_L}\right) \Big\}  \cr
\noalign{\kern 5pt}
&&\hskip 0.cm
+\bar{b}_L\gamma^\mu b_L(x)
\Big\{\frac{1}{2}h_{Z^{(n)}}^L
    \left(f_{X_L}f_{X_L}-f_{b_L}f_{b_L}-f_{D_L}f_{D_L}\right)  \cr
\noalign{\kern 5pt}
&&\hskip 0.cm
+ \frac{1}{2}h_{Z^{(n)}}^R \left(f_{D_L}f_{D_L}-f_{X_L}f_{X_L}-f_{b_L}f_{b_L}\right)
+ \hat{h}_{Z^{(n)}} \left(f_{D_L}f_{b'_L}-f_{X_L}f_{b'_L}\right)  \cr
\noalign{\kern 5pt}
&&\hskip 0.cm
+ \frac{g_B}{3g_A} h_{Z^{(n)}}^B  \left(2f_{b_L}f_{b_L}-2f_{D_L}f_{D_L}
      -2f_{b'_L}f_{b'_L}-f_{X_L}f_{X_L}\right)  \Big\} \bigg] \cr
\noalign{\kern 10pt}
&&\hskip 0.cm
+ ( L \go R) ~.
\label{KKZcoupling1}
\end{eqnarray}
Similarly the couplings between $Z^{(n)}$ and leptons are given by
\begin{eqnarray}
&&\hskip -1.cm
 Z^{(n)}_\mu(x)  \, g_w\sqrt{L} \int_{1}^{z_L}dz 
 \bigg [\bar{\nu}_{\tau L}\gamma^\mu \nu_{\tau L}(x)
\bigg\{\frac{1}{2}h_{Z^{(n)}}^L
\left(f_{\nu_{\tau L}}f_{\nu_{\tau L}}+f_{L_{3X,L}}f_{L_{3X,L}}-f_{L_{2Y,L}}f_{L_{2Y,L}}\right) \cr
\noalign{\kern 5pt}
&&\hskip -.5cm
+ \frac{1}{2}h_{Z^{(n)}}^R \left(f_{\nu_{\tau L}}f_{\nu_{\tau L}}-f_{L_{3X,L}}f_{L_{3X,L}}
   +f_{L_{2Y,L}}f_{L_{2Y,L}}\right)
+ \hat{h}_{Z^{(n)}} \left(f_{L_{2Y,L}}f_{\nu '_{\tau L}}
 +f_{L_{3X,L}}f_{\nu '_{\tau L}}\right) \cr
\noalign{\kern 5pt}
&&\hskip -.5cm
- \frac{g_B}{g_A} h_{Z^{(n)}}^B f_{\nu_{\tau L}}f_{\nu_{\tau L}} \bigg\}
+\bar{\tau}_L\gamma^\mu \tau_L(x)
  \bigg\{\frac{1}{2}h_{Z^{(n)}}^L
\left(f_{L_{1X,L}}f_{L_{1X,L}}-f_{\tau_L}f_{\tau_L}-f_{L_{3Y,L}}f_{L_{3Y,L}}\right) \cr
\noalign{\kern 5pt}
&&\hskip -.5cm
+ \frac{1}{2}h_{Z^{(n)}}^R \left(f_{\tau_L}f_{\tau_L}-f_{L_{1X,L}}f_{L_{1X,L}}
  -f_{L_{3Y,L}}f_{L_{3Y,L}}\right)
+ \hat{h}_{Z^{(n)}} \left(f_{\tau_L}f_{\tau'_L}+f_{L_{1X,L}}f_{\tau'_L}\right)  \cr
\noalign{\kern 5pt}
&&\hskip -.5cm
- \frac{g_B}{g_A} h_{Z^{(n)}}^B  \left(f_{\tau_L}f_{\tau_L}+f_{L_{1X,L}}f_{L_{1X,L}}
      +f_{\tau'_L}f_{\tau'_L}\right) \bigg\} \bigg] + (L \go R) ~.
\label{KKZcoupling2}
\end{eqnarray}

\subsection{$Z_R\bar{f}f$ couplings}

The couplings between $Z_R^{(n)}$ and  quarks are given by
\begin{eqnarray}
&&\hskip -1.cm
Z^{(n)}_{R\mu}(x) \, g_w\sqrt{L}  \int_{1}^{z_L}dz  
\bigg[\bar{t}_L\gamma^\mu t_L(x)
 \bigg\{\frac{1}{2}h_{Z^{(n)}_R}^L
    \left(f_{t_L}f_{t_L}+f_{U_L}f_{U_L}-f_{B_L}f_{B_L}\right) \cr
\noalign{\kern 5pt}
&&\hskip -.5cm
 + \frac{1}{2}h_{Z^{(n)}_R}^R \left(f_{B_L}f_{B_L}+f_{U_L}f_{U_L}-f_{t_L}f_{t_L}\right) \cr
\noalign{\kern 5pt}
&&\hskip -.5cm
+ \frac{g_B}{3g_A} h_{Z^{(n)}_R}^B  \left(2f_{B_L}f_{B_L}+2f_{t_L}f_{t_L}
      +2f_{t'_L}f_{t'_L}-f_{U_L}f_{U_L} \right) \bigg\} \cr
\noalign{\kern 5pt}
&&\hskip -.5cm
+\bar{b}_L\gamma^\mu b_L(x)
\bigg\{\frac{1}{2}h_{Z^{(n)}_R}^L
    \left(f_{X_L}f_{X_L}-f_{b_L}f_{b_L}-f_{D_L}f_{D_L}\right)
 + \frac{1}{2}h_{Z^{(n)}_R}^R \left(f_{D_L}f_{D_L}-f_{X_L}f_{X_L}-f_{b_L}f_{b_L}\right) \cr
\noalign{\kern 5pt}
&&\hskip -.5cm
 + \frac{g_B}{3g_A} h_{Z^{(n)}_R}^B  \left(2f_{b_L}f_{b_L}-f_{D_L}f_{D_L}
      -f_{b'_L}f_{b'_L}-f_{X_L}f_{X_L}\right) \bigg\} \bigg] + (L \go R) ~.
\label{KKZRcoupling1}
\end{eqnarray}
The couplings between $Z_R^{(n)}$ and  leptons are  
\begin{eqnarray}
&&\hskip -1.cm
Z^{(n)}_{R\mu}(x) \, g_w\sqrt{L}  \int_{1}^{z_L}dz  
 \bigg[\bar{\nu}_{\tau L}\gamma^\mu \nu_{\tau L}(x)
 \bigg\{\frac{1}{2}h_{Z^{(n)}}^L
    \left(f_{\nu_{\tau L}}f_{\nu_{\tau L}}+f_{L_{3X,L}}f_{L_{3X,L}}-f_{L_{2Y,L}}f_{L_{2Y,L}}\right) \cr
\noalign{\kern 5pt}
&&\hskip -.5cm
 + \frac{1}{2}h_{Z^{(n)}}^R \left(f_{\nu_{\tau L}}f_{\nu_{\tau L}}-f_{L_{3X,L}}f_{L_{3X,L}}
   +f_{L_{2Y,L}}f_{L_{2Y,L}}\right)
 + \hat{h}_{Z^{(n)}} \left(f_{L_{2Y,L}}f_{\nu '_{\tau L}}
 +f_{L_{3X,L}}f_{\nu '_{\tau L}}\right) \cr
\noalign{\kern 5pt}
&&\hskip -.5cm
- \frac{g_B}{g_A} h_{Z^{(n)}}^B f_{\nu_{\tau L}}f_{\nu_{\tau L}} \bigg\}
 +\bar{\tau}_L\gamma^\mu \tau_L(x)
\bigg\{\frac{1}{2}h_{Z^{(n)}}^L
    \left(f_{L_{1X,L}}f_{L_{1X,L}}-f_{\tau_L}f_{\tau_L}-f_{L_{3Y,L}}f_{L_{3Y,L}}\right) \cr
\noalign{\kern 5pt}
&&\hskip -.5cm
 + \frac{1}{2}h_{Z^{(n)}}^R \left(f_{\tau_L}f_{\tau_L}-f_{L_{1X,L}}f_{L_{1X,L}}
  -f_{L_{3Y,L}}f_{L_{3Y,L}}\right)
 + \hat{h}_{Z^{(n)}} \left(f_{\tau_L}f_{\tau'_L}+f_{L_{1X,L}}f_{\tau'_L}\right) \cr
\noalign{\kern 5pt}
&&\hskip -.5cm
- \frac{g_B}{g_A} h_{Z^{(n)}}^B  \left(f_{\tau_L}f_{\tau_L}+f_{L_{1X,L}}f_{L_{1X,L}}
      +f_{\tau'_L}f_{\tau'_L}\right)  \bigg\}\bigg]
+ ( L\rightarrow R ) ~.,
\label{KKZRcoupling2}
\end{eqnarray}

\subsection{$A^{\hat{4}}\bar{f}f$ couplings}

All couplings of quarks and leptons to $A^{\hat 4 (n)}$ vanish.

\subsection{$W\bar{f}f$ couplings}

The coupling of quarks and leptons to  $W^{(n)-}$  are  given by
\begin{eqnarray}
&&\hskip -1.cm
W^{(n)-}_\mu(x) \, \frac{g_w\sqrt{L}}{\sqrt{2}}  \int_{1}^{z_L}dz  
\bigg[ \bar{b}_L\gamma^\mu t_L(x)
\bigg\{h_{W^{(n)}}^L\left(f_{b_L}f_{t_L}+f_{D_L}f_{U_L}\right) \cr
\noalign{\kern 5pt}
&&\hskip -.5cm
+ h_{W^{(n)}}^R \left(f_{b_L}f_{B_L}+f_{U_L}f_{X_L}\right)
+ \hat{h}_{W^{(n)}} \left(f_{b_L}f_{t'_L}-f_{b'_L}f_{U_L}\right) \bigg\} \cr
\noalign{\kern 5pt}
&&\hskip -.5cm
\bar{\tau}_L\gamma^\mu \nu_{\tau L} (x)
\bigg\{h_{W^{(n)}}^L\left(f_{\tau_L}f_{\nu_{\tau L}}+f_{L_{3X, L}}f_{L_{3Y, L}}\right) 
 + h_{W^{(n)}}^R \left(f_{L_{1X, L}}f_{\nu_{\tau L}}+f_{L_{2Y, L}}f_{L_{3Y, L}}\right) \cr
\noalign{\kern 5pt}
&&\hskip -.5cm
 + \hat{h}_{W^{(n)}} \left(f_{L_{3Y, L}}f_{\nu '_{\tau L}}-f_{\tau'_L}f_{\nu_{\tau L}}\right)
\bigg\} \bigg]
+ (L \go R) ~.
\label{KKWcoupling1}
\end{eqnarray}
The couplings of $W^{(n)+}$ are given by the Hermitian conjugate of (\ref{KKWcoupling1}).

\subsection{$W_R\bar{f}f$ couplings}

The couplings between $W_R^{(n)-}$ and  quarks are given by
\begin{eqnarray}
&&\hskip -1.cm
W^{(n)-}_{R\mu}(x) \, \frac{g_w\sqrt{L}}{\sqrt{2}}  \int_{1}^{z_L}dz  
\bar{b}_L\gamma^\mu t_L(x)\Big\{
h_{W_R^{(n)}}^L \left(f_{b_L}f_{t_L}+f_{D_L}f_{U_L}\right)
 +h_{W_R^{(n)}}^R \left(f_{b_L}f_{B_L}+f_{U_L}f_{X_L}\right) \Big\} \cr
\noalign{\kern 10pt}
&&\hskip -.cm
+ ( L \go R) = 0 ~.
\label{KKWRcoupling1}
\end{eqnarray}
In the last equality the use of  the explicit form for wave functions has been made.
Similarly for leptons the couplings are
\begin{eqnarray}
&&\hskip -1.cm
W^{(n)-}_{R\mu}(x)  \, \frac{g_w\sqrt{L}}{\sqrt{2}}  \int_{1}^{z_L}dz
\bar{\tau}_L\gamma^\mu \nu_{\tau L}(x)
\Big\{ h_{W_R^{(n)}}^L \left(f_{\tau_L}f_{\nu_{\tau L}}
 +f_{L_{3X, L}}f_{L_{3Y, L}}\right) \cr
\noalign{\kern 10pt}
&&\hskip -.cm
 +h_{W_R^{(n)}}^R \left(f_{\nu_{\tau L}}f_{L_{1X, L}}
  +f_{L_{2Y, L}}f_{L_{3Y, L}}\right) \Big\} + (L \go R) = 0 ~.
\label{KKWRcoupling2}
\end{eqnarray}
In other words, the couplings between $W_R^{(n)}$ and quarks or leptons  vanish.


\section{HELICITY AMPLITUDES}
Here we provide formulas useful for calculations of 
cross sections discussed in this paper. 
We begin with the following interaction between 
a massive gauge boson ($A_\mu$) with mass $m_A$ and 
a pair of the SM fermions, 
\begin{eqnarray}
 {\cal L}_{\rm int}= J^\mu A_{\mu}
 =\bar{f} \gamma^\mu(g_{f_L}^A P_L+g_{f_R}^A P_R)f A_{\mu} .
\end{eqnarray}
A helicity amplitude for the process 
$f(\alpha)\bar{f}(\beta)\rightarrow F(\delta)\bar{F}(\gamma)$ 
is given by
\begin{eqnarray}
 {\cal M}(\alpha,\beta;\gamma,\delta)=
 \frac{g_{\mu \nu}}{s - m_A^2 + i m_A \Gamma_A}
J_{\rm in}^\mu(\alpha, \beta) J_{\rm out}^\nu(\gamma, \delta),
\end{eqnarray}
where $\alpha, \beta$ ($\gamma,\delta$) denote initial (final) 
spin states for fermion and antifermion, respectively, and 
$\Gamma_A $ is the total decay width of the $A$ boson. 
We have used the 't Hooft--Feynman gauge for the gauge boson propagator 
and there is no contribution from Nambu-Goldstone modes 
in the process with the massless initial states.  

The currents for initial and final states 
are explicitly given by
\begin{eqnarray}
 J_{\rm in}^\mu(+,-)=-\sqrt{s}g_{f_R}^A(0,1,i,0)\,,
\nonumber\\
 J_{\rm in}^\mu(-,+)=-\sqrt{s}g_{f_L}^A(0,1,-i,0)\,,
\end{eqnarray}
and
\begin{eqnarray}
 J_{\rm
  out}^{\mu}(+,-)&=&-\sqrt{s}g_{F_R}^A(0,\cos\theta,-i,-\sin\theta), 
\nonumber \\
 J_{\rm
  out}^{\mu}(-,+)&=&\sqrt{s}g_{F_L}^A(0,-\cos\theta,-i,\sin\theta)\,,
\end{eqnarray}
where 
 $\theta$ is the scattering angle 
 and $f$($F$) denotes a flavor of the initial (final) state of fermions.

\end{appendix}

\vskip 1cm


\leftline{\large \bf References}
\vskip 10pt

\renewenvironment{thebibliography}[1]
         {\begin{list}{[$\,$\arabic{enumi}$\,$]}  
         {\usecounter{enumi}\setlength{\parsep}{0pt}
          \setlength{\itemsep}{0pt}  \renewcommand{\baselinestretch}{1.2}
          \settowidth
         {\labelwidth}{#1 ~ ~}\sloppy}}{\end{list}}

\def\jnl#1#2#3#4{{#1}{\bf #2},  #3 (#4)}

\def\Zphys{{\em Z.\ Phys.} }
\def\jssc{{\em J.\ Solid State Chem.\ }}
\def\jpsJ{{\em J.\ Phys.\ Soc.\ Japan }}
\def\ptps{{\em Prog.\ Theoret.\ Phys.\ Suppl.\ }}
\def\PTP{{\em Prog.\ Theoret.\ Phys.\  }}
\def\JMP{{\em J. Math.\ Phys.} }
\def\NPB{{\em Nucl.\ Phys.} B}
\def\NP{{\em Nucl.\ Phys.} }
\def\PLB{{\it Phys.\ Lett.} B}
\def\PL{{\em Phys.\ Lett.} }
\def\PRL{\em Phys.\ Rev.\ Lett. }
\def\PRB{{\em Phys.\ Rev.} B}
\def\PRD{{\em Phys.\ Rev.} D}
\def\PRe{{\em Phys.\ Rep.} }
\def\AP{{\em Ann.\ Phys.\ (N.Y.)} }
\def\RMP{{\em Rev.\ Mod.\ Phys.} }
\def\ZPC{{\em Z.\ Phys.} C}
\def\SCI{\em Science}
\def\CMP{\em Comm.\ Math.\ Phys. }
\def\MPLA{{\em Mod.\ Phys.\ Lett.} A}
\def\IJMPA{{\em Int.\ J.\ Mod.\ Phys.} A}
\def\IJMPB{{\em Int.\ J.\ Mod.\ Phys.} B}
\def\EPJC{{\em Eur.\ Phys.\ J.} C}
\def\PR{{\em Phys.\ Rev.} }
\def\JHEP{{\em JHEP} }
\def\JCAP{{\em JCAP} }
\def\cmp{{\em Com.\ Math.\ Phys.}}
\def\JPA{{\em J.\  Phys.} A}
\def\JPG{{\em J.\  Phys.} G}
\def\NJP{{\em New.\ J.\  Phys.} }
\def\CQG{\em Class.\ Quant.\ Grav. }
\def\ATMP{{\em Adv.\ Theoret.\ Math.\ Phys.} }
\def\ibid{{\em ibid.} }

\renewenvironment{thebibliography}[1]
         {\begin{list}{[$\,$\arabic{enumi}$\,$]}  
         {\usecounter{enumi}\setlength{\parsep}{0pt}
          \setlength{\itemsep}{0pt}  \renewcommand{\baselinestretch}{1.2}
          \settowidth
         {\labelwidth}{#1 ~ ~}\sloppy}}{\end{list}}

\def\reftitle#1{{\it ``#1''},}    


\end{document}